\documentclass[12pt]{iopart}

\usepackage{iopams}  

\usepackage{color}      
\usepackage{graphicx}
\usepackage{dcolumn}
\usepackage{bm}
\usepackage{mdwlist}
\usepackage{textcomp}
\usepackage{hyperref}
\usepackage[T1]{fontenc}
\usepackage[latin1]{inputenc}
\usepackage{ae}
\usepackage{verbatim}
\usepackage{float}
\usepackage{epsfig}
\usepackage{pstricks,pst-grad}

\newcommand{\R}{R_{\rm eo}}                                     
\newcommand{\rg}{\rho_{\rm g}}                                  
\newcommand{\Nbar}{{\bar{\cal N}}}
\newcommand{\ie}{{\em i.e.}}
\newcommand{\eg}{{\em e.g.}}
\newcommand{\eqref}[1]{{Eq.~(\ref{#1})}}

\begin{document}

\title[Flow past hard and soft surfaces]{Molecular transport and flow past hard and soft surfaces: Computer
simulation of model systems %
\thanks{ Special issue: Microfluidics %
} }

\author{F. L{\'{e}}onforte,$^{1}$ J. Servantie,$^{1,2}$ C. Pastorino,$^{1,3}$ and M. M{\"u}ller$^{1}$}

\address{ $^{1}$ Institut f{\"u}r Theoretische Physik, Georg-August-Universit{\"a}t, 37077 G{\"o}ttingen, Germany 
\\
$^{2}$ Faculty of Engineering and Natural Sciences, Sabanci University, Orhanli 34956 Tuzla, Istanbul, Turkey 
\\
$^{3}$ Grupo de Materia Condensada, Departamento de F{\'{i}}sica, Centro At{\'{o}}mico Constituyentes (CNEA-CONICET), Av.~Gral.~Paz 1499, 1650 San Mart{\'{i}}n, Pcia.~Buenos Aires, Argentina 
}

\ead{mmueller@theorie.physik.uni-goettingen.de}

\begin{abstract}
The properties of polymer liquids on hard and soft substrates are investigated by molecular dynamics simulation of a coarse-grained bead-spring model and dynamic single-chain-in-mean-field (SCMF) simulations of a soft, coarse-grained polymer model. Hard, corrugated substrates are modelled by an FCC Lennard-Jones solid while polymer brushes are investigated as a prototypical example of a soft, deformable surface. From the molecular simulation we extract the coarse-grained parameters that characterise the equilibrium and flow properties of the liquid in contact with the substrate: the surface and interface tensions, and the parameters of the hydrodynamic boundary condition. The so-determined parameters enter a continuum description like the Stokes equation or the lubrication approximation.

At high temperatures the Navier slip condition provides an appropriate description of the flow past hard, corrugated surfaces. The position, $x_{b}$, where the hydrodynamic boundary condition is to be enforced, agrees with the location of the liquid-solid interface and the slip length can be consistently identified by comparing planar shear flow and parabolic, pressure-driven flow.  If the surface become strongly attractive or the surface is coated with a brush, the Navier slip condition will fail to consistently describe the flow at the boundary. This failure can be traced back to a boundary layer with an effective, higher viscosity.

The solvent flow past a polymer brush induces a cyclic, tumbling motion of the tethered chain molecules. The collective motion gives rise to an inversion of the flow in the vicinity of the grafting surfaces and leads to strong, non-Gaussian fluctuations of the molecular orientations in the flow. Both, molecular dynamics as well as dynamic SCMF simulations, provide evidence that the flow past a polymer brush cannot be described by Brinkmann's equation.

The hydrodynamic boundary condition is an important parameter for predicting the motion of polymer droplets on a surface under the influence of an external force. The steady state velocity is dictated by a balance between the power that is provided by the external force and the dissipation.  If there is slippage at the liquid-solid interface, the friction at the solid-liquid interface and the viscous dissipation of the flow inside the drop will be the dominant dissipation mechanisms; dissipation at the three-phase contact line appears to be less important on a hard surface.

On a soft, deformable substrate like a polymer brush, we observe a lifting up of the three-phase contact line. Controlling the grafting density and the incompatibility between the brush and the polymer liquid we can independently tune the softness of the surface and the contact angle and thereby identify the parameters to maximise the deformation at the three-phase contact. 
\end{abstract}
\noindent \textit{Keywords}: Navier boundary condition, slip length, molecular simulation, polymer brushes

\submitto{\JPCM}

\maketitle
\section{Introduction}

The equilibrium properties of a liquid on a solid substrate are chiefly determined by the balance of surface and interface tensions \cite{Young,Cahn77,deGennes1985,Sikkenk87,Vanswol91,Adams91}.  Balancing the tensions parallel to the substrate at the three-phase contact between the solid, the liquid and its coexisting vapour, Young derived in 1805 the relation \cite{Young} \begin{equation} \gamma_{{\rm LV}}\cos\Theta+\gamma_{{\rm LS}}=\gamma_{{\rm VS}}\label{Young}\end{equation} between the macroscopic contact angle, $\Theta$, of a drop, the liquid-vapour interface tension, $\gamma_{{\rm LV}}$, and the surface tensions, $\gamma_{{\rm LS}}$ and $\gamma_{{\rm VS}}$, of the liquid and its vapour in contact with the substrate. These tensions depend on the microscopic details of the liquid and the substrate, and their prediction requires molecular simulation of a microscopic model of the materials. Using the so-determined tensions, one can use Young's equation to predict the macroscopic equilibrium properties of droplets on surfaces.

%
\begin{figure}[!t]
 \includegraphics[clip,width=1\columnwidth]{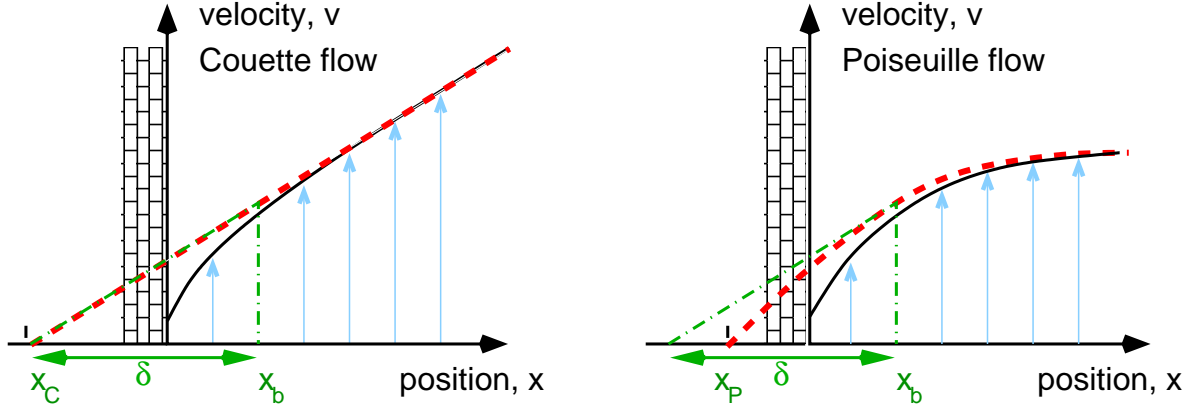}
\caption{$v_{\|}(x)$, tangential to the solid substrate. Note that the velocity of the liquid never reaches zero, \ie, there is an apparent, microscopic slip. The red dashed lines indicate the linear and parabolic hydrodynamic velocity profiles, $v_{\|}^{{\rm C,hydro}}$ and $v_{\|}^{{\rm P,hydro}}$, according to \eqref{CouetteH} and \eqref{PoiseilleH} for Couette and Poiseuille flow, respectively. These profiles are obtained by fitting the measured velocity profiles, $v_{\|}(x)$, far away from the solid substrate, to the asymptotic hydrodynamic continuum prediction.  The extrapolation of these hydrodynamic velocity profiles to zero mark the positions, $x_{{\rm C}}$ and $x_{{\rm P}}$. The green triangle indicates the Navier-slip condition, \eqref{Navier}, and the hydrodynamic position, $x_{{\rm b}}$, and slip length, $\delta$, are indicated.}
\label{fig1} 
\end{figure}

The flow of liquids past surfaces is of paramount importance in engineering applications \cite{Ellis04,Quake2005,Eijkel07,Lauga07,Bocquet07}.  On a macroscopic scale it is often assumed that the liquid velocity at a boundary equals the velocity of the surface. On a microscopic scale, however, this no-slip condition can be violated and the molecules of the liquid may slip past the surface. This slippage at the solid-liquid interface and the concomitant details of the velocity profile at the solid-liquid boundary have attracted abiding interest. Balancing the viscous stress due to the velocity gradient inside the fluid with shear viscosity, $\eta$, against the friction stress of the liquid slipping past the solid substrate, Navier formulated in 1823 a hydrodynamic boundary condition \cite{Navier1823}

\begin{equation}
\underbrace{\lambda v_{\|}\Bigr|_{x_{{\rm b}}}}_{{\rm friction\, stress\, at\, solid}}=\underbrace{\eta\left.\frac{\partial v_{\|}}{\partial x}\right|_{x_{{\rm b}}}}_{{\rm viscous\, stress\, in\, fluid}}\label{Navier}
\end{equation}
where $\lambda$ is a friction coefficient that quantifies the linear relation between the tangential velocity at the surface $v_{\|}|_{x_{{\rm b}}}$ and the friction stress. $x_{{\rm b}}$ denotes the position, at which this Navier-slip boundary condition is to be applied. The ratio $\delta=\eta/\lambda$ is a length and is denoted as slip length as illustrated in figure \ref{fig1}. This condition serves as a boundary condition to a continuum description of the fluid flow by a Navier-Stokes equation. Molecular simulations can measure the viscosity, $\eta$, of the bulk liquid and the two parameters -- the slip length, $\delta$, and the position, $x_{{\rm b}}$, of the hydrodynamic boundary condition -- of the Navier-slip condition and thereby transfer information from a microscopic, particle-based description to a continuum one.

In the following we will utilise molecular simulations of coarse-grained polymer models to extract the parameters that relate the flow properties of a particle-based model to a macroscopic description and test the validity of the assumptions of the macroscopic description on small length scales. In the next section we will introduce the particle-based models and simulation techniques. Then, we investigate the flow past hard, corrugated substrates and soft, penetrable, brush-coated substrates.  We find that the routinely used Navier-slip boundary condition may fail to describe the near-surface velocity profiles independent from the type of flow \cite{Muller08,Servantie08b,Pastorino09} and rationalise this observation by a simple, phenomenological model \cite{Servantie08b}.  Moreover, we observe the reversal of the flow direction in the vicinity of the grafting surface of a polymer brush in flow \cite{Muller08}.  The fourth section studies the behaviour of polymer drops on hard and soft surfaces and illustrates that the slip at the solid-liquid boundary is an important dissipation mechanism for the motion of small drops \cite{Servantie08}. The manuscript concludes with a brief outlook on dewetting of thin liquid films and perspectives to future molecular simulations.

\section{Models and Methods}

The surface and interface tensions as well as the viscosity and friction of the motion of simple liquids at surfaces ultimately depend on the molecular structure on the atomistic scale. In polymer systems, however, the extended structure of the molecules imparts some universal aspects onto the behaviour of these macromolecules at surfaces and interfaces.  Varying the molecular weight and, in case of brush-coated surfaces, the grafting density one can experimentally control the wettability, adhesion, and friction of a surface and the shear viscosity without altering the interactions on the atomistic scale. Therefore, polymeric systems are often described by coarse-grained models. In the following, we will use two distinct computational approaches -- molecular dynamics simulation of a Lennard-Jones, bead-spring model \cite{Kremer90,LJ_pot2,Muller00i} and dynamic single-chain-in-mean-field (SCMF) simulations of a soft, coarse-grained model of a dense polymer melt \cite{Muller05,Daoulas06b,Detcheverry09}.

\subsection{Molecular dynamics of a Lennard-Jones bead-spring model}

Molecular dynamics simulations are performed using a widely used coarse-grained polymer model \cite{Kremer90}. A macromolecule is treated as a string of $N$ beads, connected by nonlinear springs to form a linear chain molecule. The beads interact with a Lennard-Jones potential

\begin{equation}
U_{{\rm nb}}^{\alpha\beta}(r)=4\epsilon_{\alpha\beta}\Bigl[\left(\frac{\sigma_{\alpha\beta}}{r}\right)^{12}-\left(\frac{\sigma_{\alpha\beta}}{r}\right)^{6}\Bigr]\label{LJpot}
\end{equation}
 For particle distances, $r$, larger than a cut-off distance, $r_{c}$, the potential is cut-off and shifted such that it is continuous at $r_{c}$. $r_{c}=2^{1/6}\sigma_{\alpha\beta}$ gives rise to a purely repulsive interaction \cite{Kremer90}, which is appropriate to simulate dense polymer liquids. In order to study polymer droplets with a free surface, the liquid must coexist with its vapour and, to this end, the non-bonded interactions must include a longer-ranged, attractive contribution. This attraction can be included by choosing $r_{c}=2\cdot2^{1/6}\sigma_{\alpha\beta}$ \cite{LJ_pot2,Muller00i,Kreer04} or $r_{c}=2.5\sigma_{\alpha\beta}$ \cite{Grest99}.  Indices, $\alpha,\beta$, are related to different bead species, \ie, segments of the polymer liquid, segments of the polymer brush, or particles of the substrate. $\epsilon$ and $\sigma$ set the energy and length scales of the Lennard-Jones model. Adjacent segments along the chain molecule are connected by an anharmonic, finite extensible, nonlinear elastic (FENE) potential \cite{Kremer90},

\begin{equation}
U_{{\rm b}}(r)=-\frac{1}{2}kR_{{\rm o}}^{2}\ln{\left[1-\left(\frac{r}{R_{{\rm o}}}\right)^{2}\right]}\label{FENEpot}
\end{equation}
 with $k=30\epsilon/\sigma^{2}$ and $R_{{\rm o}}=1.5\sigma$. These parameters are chosen so that unphysical bond crossings and chain breaking are eliminated. In the following, all quantities will be expressed in terms of molecular diameter $\sigma\equiv1$, binding energy $\epsilon\equiv1$ and characteristic time $\tau\equiv\sqrt{m\sigma^{2}/\epsilon}$.

The equations of motion are solved via the velocity-Verlet algorithm.  The temperature is kept constant using a dissipative particle dynamics (DPD) thermostat \cite{DPD1,DPD4b,DPD2b,DPD3b,Pastorino07}. This thermostat conserves the local momentum and thus results in hydrodynamic flow behaviour on large length and long time scales. It adds to the total conservative force, \eqref{LJpot} and \eqref{FENEpot}, a dissipative force, $\textbf{F}_{i}^{D}$, and a random force, $\textbf{F}_{i}^{R}$.  Both forces are applied in a pairwise manner, such that the sum of thermostatting forces acting on a particle pair vanishes. Let $\Gamma$ be the friction constant, the dissipative force takes the form 

\begin{equation}
\textbf{F}_{i}^{D}=-\Gamma\sum_{j\neq i}\omega^{D}(r_{ij})\left(\hat{\textbf{r}}_{ij}.\textbf{v}_{ij}\right)\hat{\textbf{r}}_{ij}\label{Fdiss}
\end{equation}
 where $\hat{\textbf{r}}_{ij}=(\textbf{r}_{i}-\textbf{r}_{j})/r_{ij}$ and $\textbf{v}_{ij}=\textbf{v}_{i}-\textbf{v}_{j}$. We choose the weight functions:

\begin{equation}
\omega^{D}(r_{ij})=\left\{ \begin{array}{ll}
\left(1-r_{ij}/r_{c}\right)^{2} & \mbox{\ensuremath{\, r_{ij}<r_{c}}}\\
0 & \mbox{\ensuremath{\, r_{ij}\geq r_{c}}}\end{array}\right.\label{Wfunction}
\end{equation}
 with $r_{c}$ identical to the one used in \eqref{LJpot}. The random force is given by:

\begin{equation}
\textbf{F}_{i}^{R}=\xi\sum_{j\neq i}\omega^{R}(r_{ij})\theta_{ij}\hat{\textbf{r}}_{ij}\label{Frandom}
\end{equation}
 where $\theta_{ij}$ is a random variable with zero mean, unit variance, second moment $\langle\theta_{ij}(t)\theta_{kl}(t')\rangle=\left(\delta_{ij}\delta_{jl}+\delta_{il}\delta_{jk}\right)\delta(t-t')$, and $\theta_{ij}=\theta_{ji}$. The weight functions, $\omega^{R}(r_{ij})$, satisfy the fluctuation-dissipation theorem, $\left[\omega^{R}\right]^{2}=\omega^{D}$.  Friction, $\Gamma$, and noise strength $\xi$ define the temperature via $\xi^{2}=2k_{B}T\Gamma$. We choose $\Gamma=0.5/\tau$ in all our simulations. The equation of motions are integrated with a time step of $\Delta t/\tau=0.002$ or $0.005$. The simulations are performed using a parallel molecular dynamics program based on force-decomposition \cite{PLIMPTON95}, which is particularly suitable for spatially inhomogeneous systems like droplets, or LAMMPS \cite{LAMMPS}, which employs a geometric parallelisation strategy.

%
\begin{table}[!t]
 \begin{tabular}{|l||c|c|c|}
\hline 
N  & $10$  & $100$  & $200$ \tabularnewline
\hline
\hline 
$\R$ {[}$\sigma${]}  & $3.57$  & $12.28$  & $15.2$ \tabularnewline
\hline 
$\eta$ {[}$\sqrt{m\epsilon}/\sigma^2${]}  & $6.4\pm0.1$  & $73\pm4$  & $120\pm6$\tabularnewline
\hline 
$6D$ {[}$\sigma^{2}/\tau${]}$\times10^{3}$  & $78$  & $3.1$  & $1.6$ \tabularnewline
\hline
\end{tabular}\caption{\label{Tab1} Bulk properties of the Lennard-Jones polymer liquid
for $k_{B}T=1.2\epsilon$, $\rho_{{\rm L}}\sigma^{3}\approx0.8$,
and $r_{c}=2.5\sigma$.}
\end{table}

Some bulk properties for cut-off $r_{c}=2.5\sigma$ are compiled in table \ref{Tab1}. The data for the model with $r_{c}=2\cdot2^{1/6}\sigma$ are similar. Two temperatures, $k_{B}T/\epsilon=1.68$ \cite{Muller03d,Muller03c,Pastorino06,Macdowell06,Pastorino09} and $1.2$ \cite{Servantie08,Servantie08b}, have been studied for $r_{c}=2\cdot2^{1/6}\sigma$ and the density of the liquid that coexists with its vapour is $\rho_{{\rm L}}\sigma^{3}=0.61$ and $0.79$, respectively.  The density of the vapour is negligible at these temperatures far below the Theta-temperature, $k_{B}\Theta/\epsilon\approx3.3$, of our model \cite{Muller00i}. The spatial extension of a chain molecule is characterised by an average, mean-squared, end-to-end polymer distance $\R^{2}=\langle R_{e}^{2}\rangle^{1/2}$. For $N=10$ we observe $\R/\sigma=3.7(1)$ and $3.6(1)$ at the higher and lower temperatures, respectively. Thus, this chain length corresponds to very small values of the invariant degree of polymerisation, $\Nbar\equiv(\rho_{{\rm L}}\R^{3}/N)^{2}=9$ and $14$.

%
\begin{figure}[!t]
 \includegraphics[clip,width=0.6\columnwidth]{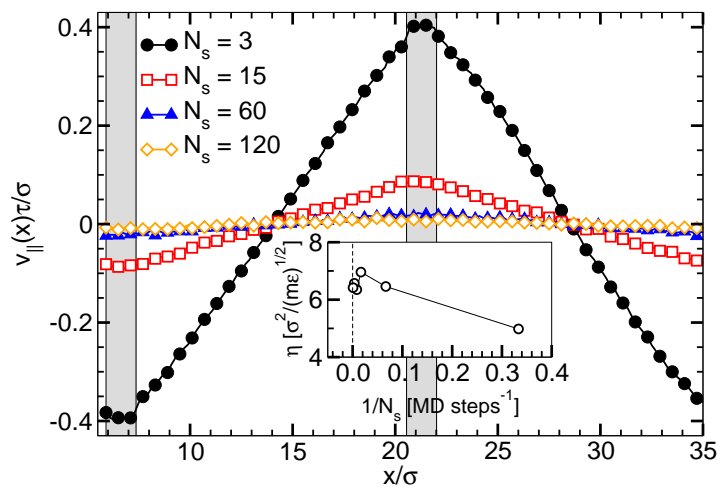}
\caption{$k_{B}T/\epsilon=1.2$ maintaining a steady-state momentum flux by exchanging of particle momenta in the shaded region according to Ref.~\cite{MullerPlathe99}.  $N_{s}$ denotes the number of molecular dynamics steps between the exchanges of momenta. The inset extrapolates the results to vanishing momentum flux. The cubic simulation cell has a spatial extent of $L=29.3\sigma$ and contained $19\,160$ Lennard-Jones segments.}
\label{figadd1} 
\end{figure}

The surface tension, $\gamma$, of the liquid-vapour interface can be measured using a slab geometry \cite{Sikkenk1988,Muller00i,Varnik00} where the interfaces are parallel to the $xy$-plane. $\gamma$ is determined by the anisotropy of the pressure

\begin{equation}
2\gamma A=V\left(\langle P_{zz}\rangle-\frac{\langle P_{xx}\rangle+\langle P_{yy}\rangle}{2}\right)
\end{equation}
 where the interfacial area is $A=L_{x}\times L_{y}$ and the factor $2$ arises because the periodic boundary conditions stabilise two liquid-vapour interfaces in the simulation box. The components of the pressure tensor are given by virial formula

\begin{equation}
P_{\alpha\beta}=\frac{1}{V}\left(\sum_{i}mv_{i\alpha}v_{i\beta}+\sum_{i<j}{F}_{ij\alpha}r_{ij\beta}\right)
\end{equation}
 where the sum over $i$ and $i<j$ runs over all particles or pairs of particles, respectively. $m$ is the mass of a segment, $v$ its velocity and ${\bf F}_{ij\alpha}$ denotes the component, $\alpha$, of the force between particles $i$ and $j$ that are a distance $r_{ij}$ apart. For $k_{B}T/\epsilon=1.2$ this procedure yields $\gamma\sigma^{2}/\epsilon=0.515(3)$ \cite{Servantie08}, while grand-canonical Monte-Carlo simulations yield the value $\gamma\sigma^{2}/\epsilon=0.16(1)$ for $k_{B}T/\epsilon=1.68$ and $r_{c}=2\cdot2^{1/6}\sigma$ \cite{Muller00i}.

The Rouse time $\tau_{{\rm R}}$ is related to the self-diffusion coefficient via $D=\R^{2}/\left(3\pi^{2}\tau_{p}\right)$. For the high temperature one finds $D=0.05\sigma^{2}/\tau$ (for $r_{c}=2\cdot2^{1/6}\sigma$), and at the lower temperature one obtains $D=0.0157(30)\sigma^{2}/\tau$ or $0.013(3)\sigma^{2}/\tau$ for $r_{c}=2\cdot2^{1/6}\sigma$ and $r_{c}=2.5\sigma$, respectively. The Rouse time characterises the longest relaxation time of the molecular conformations in equilibrium.  The product of $\tau_{{\rm R}}$ and the shear rate, $\dot{\gamma}$, defines the Weissenberg number.

The shear viscosity, $\eta$, has been computed using the reverse, non-equilibrium molecular dynamics method \cite{MullerPlathe99,Bordat02,Tenney10}.  To this end, one divides the system into slabs and exchanges the momenta between the two slabs that are separated by half the system size as indicated by the shaded regions in figure \ref{figadd1}. At each momentum swap, one exchanges the tangential momenta, $p_{y}=mv_{\|}$, of the particles with the smallest $v_{\|}$ in the centre, shaded slab with the momenta of the particle with the largest $v_{\|}$ in the left, shaded slab. This procedure imposes a momentum transfer between the slabs and results in a linear velocity profile, $v_{\|}(x)$, between the shaded regions. From the gradient of the velocity and the momentum flux, one obtains the viscosity according to \cite{MullerPlathe99}

\begin{equation}
\eta=\frac{\Sigma_{xy}}{\frac{\partial v_{\|}}{\partial x}}      \qquad\mbox{with}\quad\Sigma_{xy}=\frac{\Delta p/\Delta t}{2A}
\label{FMP}
\end{equation}
 where $\Delta p$ denotes the momenta exchange during the time interval, $\Delta t$, and $\Sigma_{xy}$ the shear stress. The number of exchange momentum swaps $N_{s}=3$ and $N_{s}=120$ were chosen in order to verify a linear velocity profile inside the fluid. Depending on $N_{s}$, we performed simulation runs of length between $5\,000\tau$ and $24\,000\tau$, while data were collected after half the simulation run, when a steady state is reached. The simulation cell was comprised of $19\,160$ particles. These results for the viscosity of the model with cut-off $r_{c}=2.5\sigma$ are very similar to the results $\eta=5.3(2)\,\sqrt{m\epsilon}/\sigma^2$ \cite{Servantie08} obtained from a Green-Kubo relation \cite{hansen-mcdonald}

\begin{equation}
\eta_{{\rm GK}}(t)=\frac{V}{k_{B}T}\int_{0}^{t}{\rm d}t'\;\left\langle P_{xy}(t')P_{xy}(0)\right\rangle \label{eq:visco_GK}
\end{equation}
 for the model with $r_{c}=2\cdot2^{1/6}\sigma$ at $k_{B}T/\epsilon=1.2$ using a molecular dynamics run of length $200\,000\tau$ for $5440$ particles \cite{Servantie08}. At the higher temperature, $k_{B}T/\epsilon=1.68$, measuring the stress in a non-equilibrium molecular dynamics simulation yields $\eta=1.9(3)\, \sqrt{m\epsilon}/\sigma^2$ \cite{Pastorino06, Pastorino09}.  In the Lennard-Jones bead-spring model the friction coefficient of the DPD thermostat is chosen small enough that its value does not have a pronounced influence on the dynamics, \ie, the segmental friction chiefly stems from the bonded and non-bonded interactions. 

\subsection{Dynamic SCMF simulations of a soft, coarse-grained polymer model}

Additionally to the standard, bead-spring model we also employed a soft, coarse-grained model of a dense polymer liquid \cite{Muller05,Daoulas06b,Detcheverry09}.  In this model, neighbouring segments along the backbone of a chain molecules are bonded together via simple, harmonic springs of the form 

\begin{equation}
U_{{\rm b}}(r)=\frac{3(N-1)k_{B}T}{2}\left(\frac{r}{\R}\right)^{2}\label{UB}
\end{equation}
 where $\R$ denotes the average end-to-end distance.

Non-bonded interactions give rise to a small compressibility of the dense polymer liquid and are described by \cite{Helfand75}

\begin{equation}
{\cal H}_{{\rm nb}}=\frac{\kappa_{{\rm o}}k_{B}T}{2}\rho_{{\rm L}}\int{\rm d}^{3}{\bf r}\;\left(\phi^{2}({\bf r})-1\right)^{2}
\end{equation}
 where $\phi({\bf r})$ denotes the normalised density of the liquid which is calculated from the explicit particle coordinates. We choose the value $\kappa_{{\rm o}}N=50$, which is sufficiently large to suppress density fluctuations on the length scale of polymer coils.  These density are efficiently calculated via a collocation grid. The quadratic form of ${\cal H}_{{\rm nb}}$ can be rewritten as a sum over pairwise interactions, $U_{{\rm nb}}$, between segments \cite{Daoulas06b}.  In contrast to the Lennard-Jones potential, however, these pairwise interactions do not exhibit a harsh repulsion at short distances but rather correspond to a soft, repulsive interactions that allows coarse-grained segments to overlap. Qualitatively, this soft, repulsive interaction between effective segments, which each represent a collection of atoms, would be the expected outcome if the interactions were derived by explicitly integrating out the atomic degrees of freedom (systematic coarse-graining) \cite{Muller-Plathe02,Reith03b}.

In a Lennard-Jones, bead-spring model the monomer density cannot significantly increase above $\rho_{{\rm L}}\sigma^{3}\sim\rho_{{\rm L}}\R^{3}/N^{3/2}\sim{\cal O}(1)$ because the fluid of segments either crystallises or becomes a glass at higher densities. In a soft, coarse-grained model, the segment density can be increased much further, $\rho_{{\rm L}}\R^{3}/N^{3/2}\sim10-100$, and thus the model allows us to assess chemically realistic values of the invariant degree of polymerisation, $\Nbar\equiv[\rho_{{\rm L}}\R^{3}/N]^{2}=[\rho_{{\rm L}}\R^{3}/N^{3/2}]^{2}N$ can be studied by molecular simulation. In section.~\ref{sec:SCMF-BRUSH} we will study molecules with $\Nbar=16\,384$, which is not feasible with a Lennard-Jones bead-spring model \cite{book10}.

The properties of the soft, coarse-grained model are studied by dynamic SCMF simulations. In SCMF simulations \cite{Daoulas06b,Daoulas06,Daoulas06c} one considers a large ensemble of explicit chain configurations that independently evolve in an external field. This field

\begin{equation}
W({\bf r})=\frac{1}{\rho_{{\rm L}}}\frac{\delta{\cal H}_{{\rm nb}}}{\delta\phi({\bf r})}
\end{equation}
 mimics the effect of the non-bonded interactions. Its gradient gives rise to a force ${\bf F}_{{\rm nb}}=-\nabla W$.

In dynamic SCMF simulations, chain segments evolve via self-consistent Brownian dynamics \cite{Doyle97b,Saphiannikova98,Narayanan04} or Smart-Monte-Carlo (SMC) moves \cite{Rossky78,allen86} with trial displacements

\begin{equation}
\Delta{\bf r}_{{\rm trial}}=\langle\bar{{\bf v}}\rangle\Delta t+\frac{{\bf F}}{\zeta_{\rm o}}\Delta t+\xi\sqrt{\frac{2k_{B}T\Delta t}{\zeta_{{\rm o}}}}
\end{equation}
 $\langle\bar{{\bf v}}({\bf r})\rangle$ denotes the hydrodynamic velocity field. ${\bf F}={\bf F}_{{\rm b}}+{\bf F}_{{\rm nb}}+{\bf F}_{{\rm ex}}$ is the force acting on a segment arising from bonded and non-bonded interactions as well as external potentials due to confining surfaces.  $\zeta_{{\rm o}}$ characterises the segmental friction, and $\xi$ is a Gaussian random number with zero mean and unit variance. Such a trial displacement, ${\bf r}\to{\bf r}$, is accepted with probability

\begin{eqnarray}
A({\bf r}\to{\bf r}') & = & \min\left(1,\exp\left[-\frac{1}{k_{B}T}\Big(E({\bf r}')-E({\bf r})\right.\right.\nonumber \\
 &  & \left.\left.+\frac{{\bf F}({\bf r}')+{\bf F}({\bf r})}{2}\{{\bf r}'-{\bf r}\}+\frac{\Delta A}{4}\left\{ {\bf F}({\bf r}')^{2}-{\bf F}({\bf r})^{2}\right\} \Big)\right]\right)
\end{eqnarray}
 where $E=\sum_{{\rm bonds}}{U}_{{\rm b}}+{\cal H}_{{\rm nb}}$ denotes the total energy of the system. The SMC-algorithm samples the equilibrium properties independent from the choice of the forces, ${\bf F}$, used to bias the proposed trial moves or the magnitude of $\Delta t$.  In the limit $\Delta t\to0$, the acceptance rate of the SMC moves tends to unity and the algorithm reproduces Brownian dynamics. One can use however much larger time steps, $\Delta t=0.08\frac{\zeta_{{\rm o}}\R^{2}}{Nk_{B}T}$ than in Brownian dynamics and still reproduce the over-damped, diffusive motion of the segments in a dense fluid \cite{MullerSL}. In this case, the self-diffusion coefficient and the shear viscosity in the bulk are close to the Rouse limit, $D_{{\rm R}}=k_{B}T/(\zeta_{{\rm o}}N)$ and $\eta_{{\rm R}}\R/(\zeta_{{\rm o}}N\sqrt{\Nbar})=1/36$, respectively.  The acceptance rate of SMC moves is about $83\%$. Chains exhibit Rouse-like dynamics, which is characteristic for unentangled melts \cite{Doi}, and the (bare) Weissenberg number is defined as Wi$_{{\rm o}}\equiv\dot{\gamma}_{{\rm o}}\zeta_{{\rm o}}N\R^{2}/(3\pi^{2}k_{B}T)$ where $\dot{\gamma}_{{\rm o}}$ denotes the shear rate imposed by moving the confining surfaces with the grafted polymer chains.

%
\begin{figure}[!t]
 \includegraphics[clip,width=0.6\columnwidth]{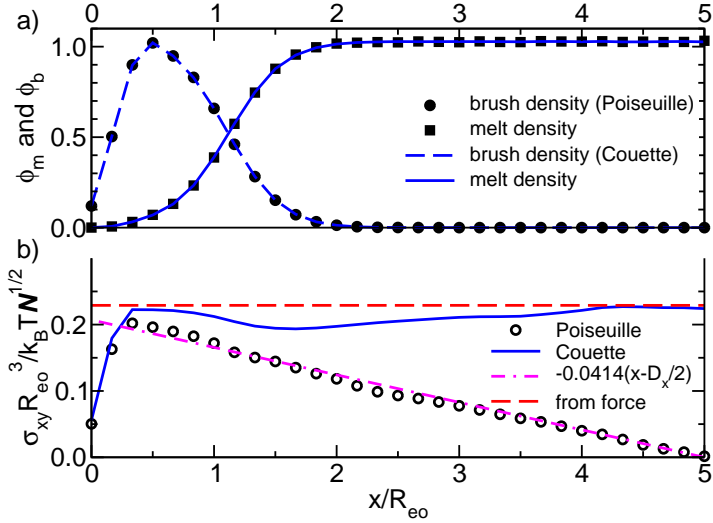}
\caption{$L_{x}=10\R$ and the lateral dimensions are $L_{y}=L_{z}=4\R$. The total normalised density, $\phi$, is comprised of the density, $\phi_{{\rm m}}$, of the melt and the density, $\phi_{{\rm b}}$, of the grafted chain molecules. Data for Poiseuille flow ($f_{\|}N=0.03k_{B}T/\R$, symbols) and Couette flow ($\dot{\gamma}_{{\rm o}}\zeta_{{\rm o}}N\R^{2}/k_{B}T=5$, Wi$_{{\rm o}}=0.247$, lines) are shown. The simulation cell contains $655\ 360$ segments. b) Kramer's stress for Poiseuille and Couette flow. The horizontal dashed line marks the stress obtained from the forces on the grafted ends for Couette flow. Adapted from Ref.~\cite{Muller08}}
\label{fig_scmf} 
\end{figure}

In the Lennard-Jones, bead-spring model, the non-bonded interactions provide an important contribution to the friction and viscosity of the liquid of monomers gives rise to corrections to the chain-length dependence of the viscosity predicted by the Rouse or reptation model. In the soft, coarse-grained model, however, the the non-bonded interactions hardly give rise to friction. Instead, the friction is chiefly generated by the random forces in case of Brownian dynamics, \ie, the parameter, $\zeta_{{\rm o}}$, of the SMC algorithm dictates the friction. Moreover, the soft non-bonded potentials in conjunction with the harmonic bonded ones allow the chains to cross through each other in the course of their motion in SCMF simulations \cite{MullerSL}. 

The {\em fluctuating, external field}, $W$, is calculated after every time step. It is important that they are frequently updated such that they mimic the instantaneous, non-bonded interactions. If this condition is fulfilled \cite{Daoulas06b} SCMF simulations will capture even subtle deviations from the Gaussian statistics in a dense polymer melt \cite{Wittmer04}. The hydrodynamic velocity, $\langle\bar{{\bf v}}\rangle$, however, represents the {\em average flow field} and must not fluctuate.  It is calculated self-consistently from the particle displacements.  First, we directly calculate an instantaneous velocity, ${\bf v}_{i}=\Delta{\bf r}_{i}/\Delta t$, of segment $i$ from its explicit displacement \cite{Narayanan06} during a SMC step (in order to retain spatial resolution) and assign it to the grid. Then, we add the instantaneous velocities of all segments, average over a time period, $T$, and normalise by the local density to obtain the average local velocity $\langle\bar{{\bf v}}({\bf r})\rangle$.  This procedure ensures that the average force, $\langle\bar{{\bf F}}\rangle$, vanishes. This time averaging procedure limits the simulation technique to stationary or slowly varying flows. For the density utilised in our simulations a time interval $T\geq800\zeta_{{\rm o}}\R^{2}/(Nk_{B}T)$ is sufficient to eliminate fluctuations of the velocity field and to yield accurate predictions in equilibrium.

In figure \ref{fig_scmf}a we present the density profile across the slit pore with width, $L_{x}=10R_{e}$. The solid surfaces are coated with a polymer brush and the tethered chains have the same physical properties as the chains in the melt. The grafting density is $\sigma_{g}\R^{2}=\sqrt{\bar{{\cal N}}}$.  Results of dynamic SCMF simulations are presented for Couette and Poiseuille flow. At this moderate grafting density, there is a broad interface between brush and melt ({}``wet brush''). In agreement with previous self-consistent Brownian dynamics studies \cite{Saphiannikova98} and molecular dynamics simulation of bead-spring models \cite{Grest99,Pastorino06}, the profile normal to the surface does not depend on flow for small shear rates. The depletion at the wall stems from a repulsive segment-wall potential $U_{{\rm wall}}(x)=\Lambda_{{\rm w}}\exp(-x^{2}/2\epsilon_{{\rm wall}}^{2})$ with $\Lambda_{{\rm w}}N=66.\bar{6}k_{B}T$ and $\epsilon_{{\rm w}}=0.15R_{e}$.  This repulsive potential impedes the Smart-Monte-Carlo algorithm to propose moves that would penetrate the hard walls and be rejected.

Measuring the forces, $F_{\|}^{{\rm graft}}$, that act on the grafted segments in Couette flow, we obtain the shear stress \begin{equation} \frac{\Sigma_{xy}\R^{3}}{k_{B}T}=\frac{F_{\|}^{{\rm graft}}\R^{3}}{L_{y}L_{z}k_{B}T}=\frac{\eta\dot{\gamma}\R^{3}}{k_{B}T}\approx0.229\sqrt{\Nbar}\end{equation} for Couette flow. This value is indicated by the horizontal dashed line in figure \ref{fig_scmf}b) and it agrees well with the prediction of the Rouse model 

\begin{equation}
\frac{\Sigma_{xy}^{{\rm Rouse}}\R^{3}}{k_{B}T\sqrt{\Nbar}}\simeq\frac{\pi^{2}}{12}{\rm Wi}_{{\rm o}}\approx0.203
\end{equation}

Alternatively, we can estimate the intramolecular stress of the melt at the centre of the film via the mean-field approximation (Kramer's formula \cite{Doi}\ie, virial of the bonded interactions, \eqref{UB}), which only accounts for the bonded contribution to the stress and disregards the contribution of all non-bonded forces.

\begin{equation}
\frac{\Sigma_{xy}^{{\rm Kramer}}\R^{3}}{k_{B}T\sqrt{\Nbar}}=(N-1)\phi\frac{\langle b_{x}b_{y}\rangle}{\R^{2}/[3(N-1)]}
\end{equation}
 where $b_{x}$ and $b_{y}$ denote the distance between bonded segments perpendicular to the surfaces and along the shear direction, respectively.  This estimate is shown in panel (b) of figure \ref{fig_scmf} for Poiseuille and Couette flow. We have utilised the same spatial assignment for the stress as for the density. Alternative assignment schemes for the stress that are locally more accurate can be envisioned \cite{Irving50}.  Thus the small variation of the stress in Couette flow may be either due to inaccuracies of the local assignment or contributions that stem from the non-bonded forces suppression density fluctuations.

The stress in Couette geometry agrees well with the result obtained from the force on the grafted segments. From the stress at the centre of the film and the velocity gradient (see figure \ref{fig5}a) we can estimate the dimensionless shear viscosity, $36\eta\R/(\zeta_{{\rm o}}N\sqrt{\bar{{\cal N}}})=1.12$ for Couette flow. The linear dependence of the stress in Poiseuille flow shown in figure \ref{fig_scmf}b is well described by $\frac{\partial\Sigma_{xy}}{\partial x}=0.0414k_{B}T\sqrt{\Nbar}/\R^{4}$ and the curvature of the parabolic velocity profile (cf.~figure \ref{fig5}a) is $\frac{\partial^{2}v_{\|}}{\partial x^{2}}=1.38k_{B}T/(\zeta_{{\rm o}}N\R^{3})$.  Both results yield 36 $\eta\R/(\zeta_{{\rm o}}N\sqrt{\bar{{\cal N}}})=1.08$.

\section{Flow past hard and soft substrates}

\subsection{Corrugated, hard substrates}
\label{sec:CHS}
%
\begin{figure}[!t]
 \includegraphics[clip,width=0.6\columnwidth]{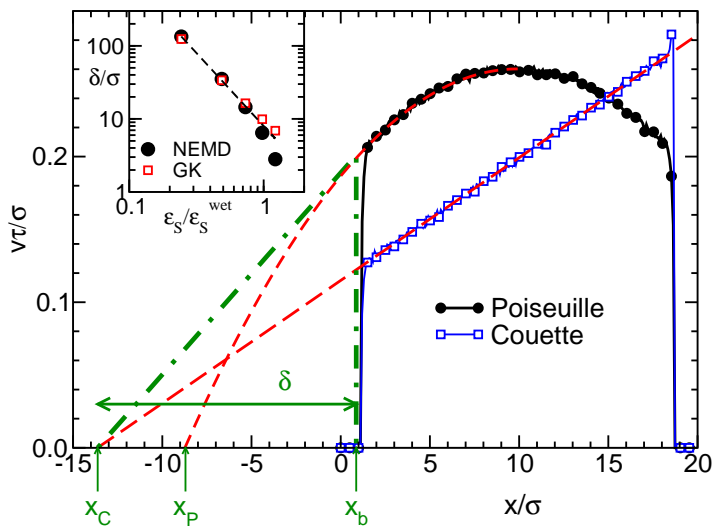}
\caption{$\epsilon_{{\rm S}}=0.6\epsilon=0.73\epsilon_{{\rm S}}^{{\rm wet}}$ and temperature, $k_{B}T/\epsilon=1.2$. The measured velocity profiles are marked by symbols, the hydrodynamic velocity profiles obtained by fitting the simulation data at the centre of the film at indicated by red, dashed lines and the construction of the slip length, $\delta$, and the hydrodynamic position is depicted by the dashed-dotted, green lines. Inset: Slip length, $\delta$, as a function of the attraction, $0.2\leq\epsilon_{{\rm S}}/\epsilon\leq1$, between solid and liquid at temperature, $k_{B}T/\epsilon=1.2$. The solid line with circles is obtained from the Couette and Poiseuille profiles using non-equilibrium molecular dynamics (NEMD) simulation, while the dashed line with squares from the Green-Kubo (GK) relation \eqref{Barrat}. The dashed line marks the behaviour $\delta\sim1/\epsilon_{{\rm S}}^{2}$ according to \eqref{Bocquet}. Adapted from Ref.~\cite{Servantie08}. }
\label{fig2} 
\end{figure}

First, we consider the flow of a Lennard-Jones, bead-spring liquid past a hard, corrugated substrate \cite{Servantie08}, which is comprised of two layers of an FCC lattice with density $\rho_{{\rm S}}\sigma^{3}=2$.  The particles of the solid substrate interact with the segments of the liquid via a Lennard-Jones potential with $\sigma_{{\rm S}}=0.75\ \sigma$ and $\epsilon_{{\rm S}}=0.2-1.0\ \epsilon$ and $r_{c}=2\cdot2^{1/6}\sigma$.  Increasing $\epsilon_{{\rm S}}$ we simultaneously alter the wettability of the surface in equilibrium as well as the friction.

The friction coefficient, $\lambda$ in \eqref{Navier} can be evaluated via a Green-Kubo relation \cite{Bocquet1993,Bocquet1994,Bocquet07}

\begin{equation}
\lambda=\frac{1}{k_{B}TA}\int_{0}^{\infty}{\rm d}t\;\left\langle F_{{\rm LS}\|}(t)F_{{\rm LS}\|}(0)\right\rangle \label{Barrat}
\end{equation}
 for the tangential forces, $F_{{\rm LS}\|}$, that the substrate exerts onto the fluid. $A$ denotes the surface area. Using the shear viscosity, $\eta$, we plot the slip length $\delta\equiv\eta/\lambda$ in the inset of figure \ref{fig2} for $k_{B}T/\epsilon=1.2$ and various values of the attraction, $\epsilon_{{\rm S}}$, between substrate and liquid. The slip length decreases as we increase the attraction between liquid and substrate, but $\delta$ remains finite at the wetting transition, which is located around $\epsilon_{{\rm S}}^{{\rm wet}}\approx0.82\epsilon$.  Qualitatively, this finding agrees with other simulation studies \cite{Thompson90b,Bocquet94,Cieplak01,Troian2004,Priezjev09}.

Barrat and Bocquet devised an approximation \cite{Barrat99b,Bocquet07}

\begin{equation}
\lambda\approx\frac{\rho_{L}S(q_{\|})}{D(q_{\|})k_{B}T}\int{\rm d}x\;\phi(x)U_{{\rm LS}}^{2}(x)\label{Bocquet}
\end{equation}
 which captures the main ingredients that qualitatively dictate the friction at the surface. $S(q_{\|})$ is the structure factor of the liquid at the surface and $q_{\|}$ is the wave vector of the surface corrugation. $D(q_{\|})$ denotes the collective diffusion coefficient at wave vector, $q_{\|}$. Assuming that the normalised density profile $\phi(z)$ is not altered by the strength $\epsilon_{{\rm S}}$ of the interaction, $U_{{\rm LS}}$, between liquid and substrate, one obtains the prediction, $\delta\sim1/\epsilon_{{\rm S}}^{2}$, which is nicely confirmed by the simulation data in figure \ref{fig2}.  Moreover, the adhesion free energy $W=\gamma_{{\rm LV}}(1+\cos\Theta)$ roughly depends linearly on $\epsilon_{{\rm S}}$ for $\epsilon_{{\rm S}}>0.4\epsilon$ \cite{Servantie08}, because 
 
\begin{equation}
\frac{{\rm d}W}{{\rm d}\epsilon_{{\rm S}}}=\frac{1}{A}\frac{\langle{\cal H}_{{\rm S}}\rangle}{\epsilon_{{\rm S}}}
\end{equation}
 with ${\cal H}_{{\rm S}}$ being the energy of the interaction between the solid substrate and the liquid, is chiefly dictated by the normalised density profile, $\phi(x)$. The latter, however, only weakly depends on $\epsilon_{{\rm S}}$ provided that (i) the temperature is sufficiently low and the liquid compressibility is concomitantly low and (ii) $\epsilon_{{\rm S}}$ is sufficiently large in order to suppress the formation of a drying layer. In this regime, the data are also compatible with a behaviour of the form, $\delta\sim(1+\cos\Theta)^{-2}$ , which has recently been observed for water at hydrophobic surfaces \cite{Huang08c,Sendner09b}.

Alternatively, one can determine the two parameters -- $\delta$ and $x_{{\rm b}}$ -- of the Navier slip condition by non-equilibrium molecular dynamics. Using Couette flow, one can only extract the combination, $x_{{\rm b}}+\delta$. Note that for chemically complex surfaces or soft, deformable substrates (\eg, brush-coated surfaces or networks), the hydrodynamic position, $x_{{\rm b}}$, of the substrate in the continuum description may not be obvious, and it may depend on external control parameters (like temperature, pressure and grafting density) \cite{Pastorino06,Pastorino09}. In order to determine both parameters independently, we study Couette and Poiseuille flow. These two flow profiles are particularly simple in the sense that they are the only ones, for which the stress depends only linearly on the distance from the surfaces \cite{Muller09c}. Therefore, a non-locality of the viscosity, which is often invoked by generalised hydrodynamic descriptions \cite{Alley83,Todd08} in order to extend the continuum description to smaller length scales, has no pronounced consequences.

Couette flow in a thin film between two parallel, hard, corrugated walls is produced in non-equilibrium molecular dynamics simulations by moving the confining surfaces in opposite directions with velocity $v_{\|}=\pm\dot{\gamma}_{{\rm o}}/(2L_{x})$, where $L_{x}$ denotes the film thickness. This results in a linear velocity profile $v_{\|}^{{\rm C,hydro}}(x)$ at the centre of the film, where a hydrodynamic continuum description is valid. 

\begin{equation}
v_{\|}^{{\rm C,hydro}}(x)=\dot{\gamma}\left(x-x_{{\rm C}}\right)
\label{CouetteH}
\end{equation}
 At the surface, however, there may be deviations from the linear velocity profile as illustrated in figure \ref{fig1}. From the velocity profile at the centre one extracts the shear rate, $\dot{\gamma}$, which may deviate from the value $\dot{\gamma}_{{\rm o}}$ imposed by the moving surfaces and the position, $x_{{\rm C}}$, where the hydrodynamic velocity profile extrapolates to zero.

Poiseuille flow in a thin film is generated by applying a force, $f_{\|}$, onto all segments. This results in a parabolic velocity profile, $v_{\|}^{{\rm P,hydro}}(x)$, at the centre of the film, which takes the form 

\begin{equation}
v_{\|}^{{\rm P,hydro}}(x)=\frac{\rho_{{\rm L}}f_{\|}}{2\eta}\left(x-x_{{\rm P}}\right)\left(L_{x}-x_{{\rm P}}-x\right)
\label{PoiseilleH}
\end{equation}
 Provided that the film thickness, $L_{x}$, is large enough to observe bulk-like behaviour at the centre, the non-equilibrium simulation of Poiseuille flow yields the shear viscosity, $\eta$, in the bulk and the position, $x_{{\rm P}}$, where the hydrodynamic velocity profile extrapolates to zero. These extrapolations are presented in figure \ref{fig2} for $\epsilon_{{\rm S}}=0.6\epsilon$.

Since the Navier-slip condition provides a boundary condition to the hydrodynamic continuum description one should use the extrapolations of the velocity profiles at the centre to extract the slip length, $\delta$, and hydrodynamic position, $x_{{\rm b}}$. These parameters are simply related to $x_{{\rm C}}$ and $x_{{\rm P}}$ via \cite{Muller08,Servantie08b}

\begin{eqnarray}
\delta & = & \sqrt{(x_{{\rm C}}-x_{{\rm P}})(L_{x}-x_{{\rm P}}-x_{{\rm C}})}\label{delta}\\
x_{{\rm b}} & = & x_{{\rm C}}+\delta\label{xb}
\end{eqnarray}

The results of these non-equilibrium molecular dynamics simulations are also plotted in the inset of figure \ref{fig2}. Gratifyingly they agree with the data extracted from the Green-Kubo relation for small and moderate attraction between polymer and substrate. For strong attraction, $\epsilon_{s}$, the Green-Kubo method however yields slightly larger values of the slip length than the non-equilibrium method. This deviation is partially due to the fact that we employ the bulk value of the shear viscosity while the structure and dynamics of the liquid is altered by the strong interactions with the substrate \cite{Goel08}.

%
\begin{figure}[!t]
 \includegraphics[clip,width=0.6\columnwidth]{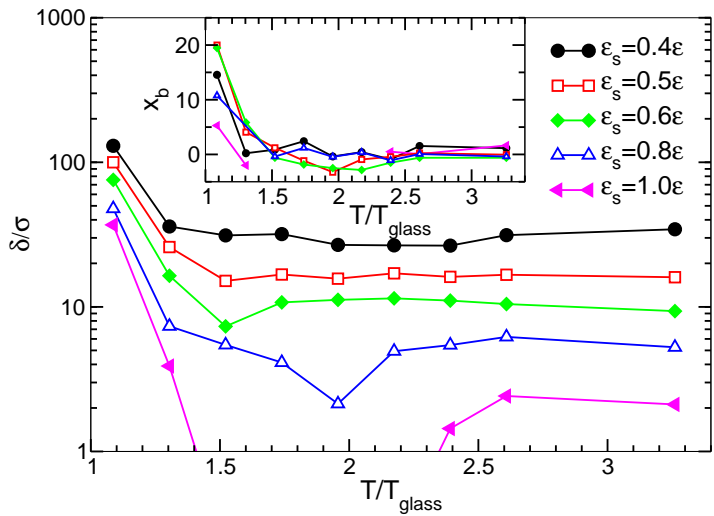}
\caption{$\delta$, versus temperature for different strengths, $\epsilon_{{\rm S}}$, of solid-fluid interaction. The inset represents the position of the hydrodynamic boundary, $x_{{\rm b}}$, versus temperature. All lengths, $\delta$ and $x_{{\rm b}}$, are measured in units of the Lennard-Jones parameter, $\sigma_{{\rm LJ}}$, and temperature, $T$, in units of the glass transition temperature, $k_BT_{\rm glass} = 0.41 \epsilon$. The thickness of the film is varied between $8\leq D/\R\leq17$, the typical velocity of the wall in Couette flow is $v_{s}=0.2\sigma/\tau$ and the typical volume force applied to generate Poiseuille flow is $0.001\leq f_{\|}\sigma/\epsilon\leq0.005$.  Adapted from Ref.~\cite{Servantie08}}
\label{fig3} 
\end{figure}

The temperature dependence of the slip length, $\delta$, and the hydrodynamic position, $x_{{\rm b}}$, extracted from non-equilibrium molecular dynamics simulations are presented in figure \ref{fig3}.  Data sets correspond to various strengths, $\epsilon_{{\rm S}}$, of the attraction between the solid and liquid as indicated in the key, and the liquid density corresponds to the temperature-dependent coexistence value. If the surface is only weakly attractive, $\epsilon_{{\rm S}}=0.4\epsilon$, the slip length, $\delta$, and the hydrodynamic position, $x_{{\rm b}}$, exhibit only a very weak temperature dependence. At low temperatures, however, when we approach the glass transition of the model \cite{Baschnagel05,Peter06} at around $k_{B}T_{{\rm glass}}/\epsilon\approx0.41$, the slip length, $\delta$, increases.  The growth of $\delta=\eta/\lambda$ for $T\to T_{{\rm glass}}$ indicates that the friction inside the liquid, $\eta$, diverges more rapidly than the friction, $\eta$, of the liquid with the corrugated substrate, whose structure is independent of $T$. For $T<T_{{\rm glass}}$, the polymer systems will become a glass solid and will move like a solid body (plug flow), which corresponds to an infinite value of the slip length. Also the hydrodynamic position, $x_{{\rm b}}$, increases slightly for $T\to T_{{\rm glass}}$ due to the increase of liquid structuring (packing) at the substrate upon cooling the system.

Already at $k_{B}T/\epsilon=0.5$, \ie, about $20\%$ above the glass transition temperature of our model, $\delta$ has increased by an order of magnitude compared to the approximately constant value at high temperature. This observation offers an explanation for the surprisingly large slip length observed in the dewetting experiments of Fetzer {\em et al.} \cite{Fetzer05,Fetzer06,Fetzer07c,Fetzer07d}, which were performed in the vicinity of the glass transition temperature.

If the attraction between solid and liquid becomes larger, we observe that the slip length, $\delta$, will first decrease upon cooling, pass through a minimum, and eventually it will diverge upon approaching $T_{{\rm glass}}$ from above for the reasons explained above. The decrease of $\delta$ upon cooling from large $T$ can be rationalised by the changes of the liquid structure in the vicinity of the solid substrate. Upon cooling, the liquid becomes more structured and this gives rise to an increased, effective viscosity in the vicinity of the surface \cite{Servantie08b}. Qualitatively, decreasing the temperature at constant $\epsilon_{{\rm S}}$ is similar to increasing the attraction, $\epsilon_{{\rm S}}$, at constant temperature; both effects lead to a more pronounced structuring of the liquid at the surface and a smaller slip length.

At large attraction between solid and liquid, we find that $x_{{\rm P}}>x_{{\rm C}}$ and \eqref{delta} has no solution. The individual velocity profiles observed for Couette and Poiseuille flow in the non-equilibrium simulations do not exhibit any unusual features compared to the situation, where a slip length can be extracted. This observation marks the failure of the Navier-slip condition, \eqref{Navier}, to consistently describe both types of flows with the same set of parameters, $\delta$ and $x_{{\rm b}}$. In this situation, \eqref{Navier} cannot be utilised as a boundary condition for a continuum description because a boundary condition should parameterise the microscopic phenomena at the boundary independently from the type and strength of the hydrodynamic flow in the bulk.

\subsection{Failure of the Navier-slip condition: a schematic, two-layer model}

%
\begin{figure}[!t]
 \includegraphics[clip,width=0.6\columnwidth]{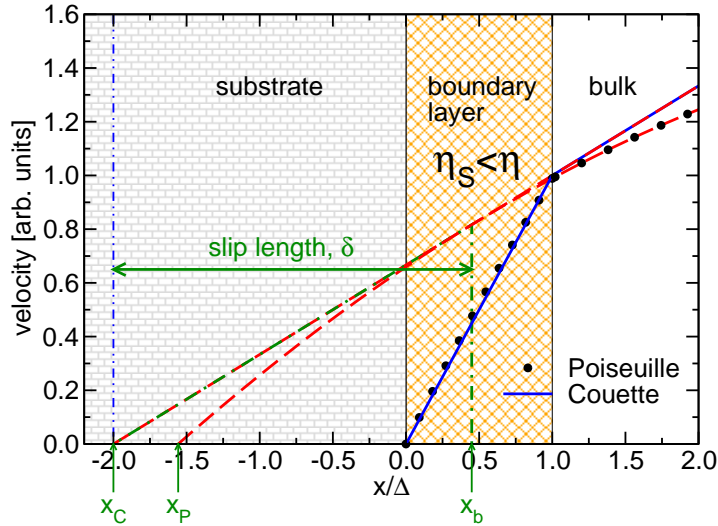}
\caption{$\delta_{{\rm S}}=0$ and the formation of a lubrication layer with $\eta_{{\rm S}}<\eta$ at the solid substrate. The velocity profiles for Poiseuille (circles) and Couette flow (black, solid line), the hydrodynamic velocity profiles at the centre and its extrapolation (red, dashed lines), and the construction of the slip length, $\delta$, and hydrodynamic position, $x_{{\rm b}}$, (green, dashed-dotted line) are sketched.}
\label{fig4} 
\end{figure}

Couette and Poiseuille flows are characterised by constant or linear stress profiles, respectively. Thus, we can rule out that the failure of the Navier-slip condition stems from a non-local relationship between velocity gradient and stress that might arise from the extended molecular conformations of the chain molecules. Nevertheless, the non-Newtonian nature of the polymer liquid may contribute to the failure because a finite driving must be applied in the non-equilibrium simulations.  Pronounced effects of shear-thinning on the slip length in polymer fluids have been observed \cite{Troian2004,Priezjev09}. Moreover, due to the spatial extension of the chain molecules, limited range of film thicknesses, $L_{x}$, accessible in the simulation may not be sufficient to observe bulk-like behaviour at the centre of the film.

Therefore, it is beneficial to rationalise the failure of the Navier-slip boundary condition in a schematic model, which qualitatively reproduces the failure observed in the non-equilibrium simulations but does not invoke the additional molecular length scales, $\sigma$ and $\R$, strictly deals with a Newtonian fluid, and can be analytically studied for all film thicknesses, $L_{x}$. In order to capture the changes of the liquid in the vicinity of the solid substrate, we consider a two-layer model, which is illustrated in figure \ref{fig4}, where we crudely approximate the gradual variation of the fluid properties as a function of the distance, $x$, from the solid surface by a boundary region of width, $\Delta$. For distances larger than $\Delta$ the fluid posses the bulk properties and it is characterised by a viscosity, $\eta$. Closer to the solid substrate, $x<\Delta$, the liquid properties deviate from the bulk and, for simplicity, we parameterise these changes by a constant effective viscosity, $\eta_{{\rm S}}$. The width is the boundary layer, $\Delta$, is the only length scale in the model.  Within the boundary layer, $0<x<\Delta$, and in the bulk, $x>\Delta$, the Newtonian liquid is described by the Navier-Stokes equation. The tangential velocities are denoted by $v_{{\rm S}}(x)$ and $v(x)$ in the boundary layer and the bulk, respectively. At the interface between the solid surface and the boundary layer, $x=0$, we impose a Navier-slip condition, \eqref{Navier}, with a microscopic slip length, $\delta_{{\rm S}}$. At the interface between the boundary layer and bulk, we require the continuity of shear stress and velocity,

\begin{equation}
\left.\eta_{{\rm S}}\frac{\partial v_{{\rm S}}}{\partial x}\right|_{x=\Delta^{-}}=\left.\eta\frac{\partial v}{\partial x}\right|_{x=\Delta^{+}}\qquad\mbox{and}\qquad\left.v_{{\rm S}}\right|_{x=\Delta^{-}}=\left.v\right|_{x=\Delta^{+}}\label{schema1}
\end{equation}
 These are the natural boundary conditions at a fluid-fluid interface.

One can straightforwardly calculate the flow profiles in the boundary region and in the bulk. For Couette flow, one obtains:

\begin{eqnarray}
v_{{\rm S}}(x) & = & \dot{\gamma}\frac{\eta}{\eta_{{\rm S}}}\left(x-\delta_{{\rm S}}\right)\qquad\mbox{for}\quad x<\Delta\\
v(x) & = & \dot{\gamma}\left(x-x_{{\rm C}}\right)\qquad\mbox{for}\quad x\geq\Delta\\
\mbox{with} &  & x_{{\rm C}}=\left(1-\frac{\eta}{\eta_{{\rm S}}}\right)\Delta-\frac{\eta}{\eta_{{\rm S}}}\delta_{{\rm S}}\nonumber 
\end{eqnarray}
 A similar calculation yields

\begin{eqnarray}
v_{{\rm S}}(x) & = & -\frac{\rho_{{\rm L}}f_{\|}}{2\eta_{{\rm S}}}\left(x^{2}-L_{x}x-D\delta_{{\rm S}}\right)\qquad\mbox{for}\quad x<\Delta\\
v(x) & = & \frac{\rho_{{\rm L}}f_{\|}}{2\eta}\left(x-x_{P}\right)\left(L_{x}-x_{{\rm P}}-x\right)\qquad\mbox{for}\quad x\geq\Delta\\
\mbox{with} &  & x_{{\rm P}}\left(L_{x}-x_{{\rm P}}\right)=-\left(1-\frac{\eta}{\eta_{{\rm S}}}\right)\Delta^{2}+L_{x}x_{{\rm C}}\nonumber 
\end{eqnarray}
 for Poiseuille flow. $L_{x}$ denotes the film thickness, which is defined by the distance between the positions where the microscopic Navier slip condition with slip length, $\delta_{{\rm S}}$, is applied.

Finally, \eqref{delta} yields the slip length

\begin{equation}
\delta=\sqrt{\Delta\frac{\eta}{\eta_{{\rm S}}}\left(\frac{\eta}{\eta_{{\rm S}}}-1\right)\left(\Delta+2\delta_{S}\right)+\left(\frac{\eta}{\eta_{{\rm S}}}\delta_{{\rm S}}\right)^{2}}\label{eqn:delta}
\end{equation}
 which is independent from the film thickness, $L_{x}$. In this schematic model, the order of magnitude of the slip length is set by the spatial extent, $\Delta$, of the boundary layer.

The first term describes the effect of the surface layer, the second term arises from the microscopic slip at the solid surface. The condition, $\eta/\eta_{{\rm S}}>1$, corresponds to the formation of a lubrication layer at the surface and results in an enhanced slip length, $\delta>\delta_{{\rm S}}$, compared to the microscopic slip at the solid-fluid interface. If, however, the solid-liquid interactions give rise to a boundary layer with a large effective viscosity, $\eta/\eta_{{\rm S}}<1$, the presence of this sticky boundary layer reduces the slip length, $\delta<\delta_{{\rm S}}$.  Moreover, if

\begin{equation}
\frac{\eta}{\eta_{{\rm S}}}\le\frac{1+2\delta_{{\rm S}}/\Delta}{\left(1+\delta_{{\rm S}}/\Delta\right)^{2}}\label{failure}
\end{equation}
 the Navier-slip condition, \eqref{Navier}, fails and $x_{{\rm P}}>x_{{\rm C}}$ as observed in the simulations of a polymer liquid in contact with a strongly attractive substrate.

Qualitatively, our schematic model can rationalise the observations in our molecular simulation: (i) At high temperature, kinetic effects will dominate the behaviour, thus $\eta_{{\rm S}}\approx\eta$. In this case, $\delta$ is equal to the microscopic slip length $\delta\approx\delta_{{\rm S}}$.  (ii) Upon cooling the fluid, the shear viscosity increases. If the solid-fluid interactions are weak, $\epsilon_{{\rm S}}<0.5\epsilon$, a lubrication layer is formed and the slip length increases, $\delta\propto(\eta/\eta_{{\rm S}})\left(\Delta+\delta_{{\rm S}}\right)$.  (iii) If the coupling between solid and liquid is strong, however, the ratio $\eta/\eta_{{\rm S}}$ decreases upon cooling and so does $\delta$. If the ratio becomes sufficiently small (cf.~\eqref{failure}), as it does in the case $\epsilon_{S}=1$ for a melt at an attractive surface, the Navier-slip condition fails.

It is an open question by what condition the Navier-slip condition, \eqref{Navier}, has to be replaced in case it fails. The schematic, two-layer model can account for the phenomenon that is observed in the molecular simulations but it introduces two additional parameters -- the effective viscosity, $\eta_{{\rm S}}$, and the thickness, $\Delta$, of the boundary layer. Moreover, the sharp interface between the boundary region and the bulk is only a crude representation of the rather gradual changes of the liquid structure and its dynamics in the vicinity of the surface. The stress balance between the viscous stress in the liquid and the friction stress at the solid substrate, which is the physical rational for \eqref{Navier}, is also part of our schematic, two-layer model because we require the stress at the interface between boundary layer and bulk to be equal, cf.~\eqref{schema1}.  Thus, one might speculate that it is the right hand side of \eqref{Navier} that relates the velocity gradient to the stress inside the fluid, which should be generalised in order to parameterise the changes of the fluid at the solid substrate. The failure of the simple, linear relation between stress and velocity gradient has often been observed for large shear rates \cite{Troian2004,Priezjev09}, where the polymer liquid exhibits shear-thinning. The schematic model, however, demonstrates that the shear rate and the concomitant non-Newtonian behaviour is not responsible for the failure and it suggests that the right hand side of \eqref{Navier} should be augmented by spatial derivatives of higher order.

\subsection{Soft, penetrable, brush-coated substrates}

\label{sec:SCMF-BRUSH}

The equilibrium and kinetic properties of the solid-liquid interface -- wettability and hydrodynamic boundary condition -- depend on the microscopic structure and dynamics of the liquid at the solid substrate.  For instance, surface coatings -- like an oxide layer -- can modify the wettability of a surface \cite{Muller01, Muller-Buschbaum05}. Molecular simulations are well suited to investigate how one can tune these properties by controlling the microscopic interactions at the contact of solid and liquid. Since we utilise a coarse-grained polymer model \cite{Kremer90}, we focus the effect of a polymer brush coating.  Polymer brushes offer a very stable and versatile strategy of tuning wettability, adhesion and friction \cite{Klein94b, Grest99}.  Moreover, the effect of a polymer brush is rather universal because it much depends on the interdigitation between the brush and melt.  The structure of the brush-melt interface is dictated by a balance of translational entropy, which favours mixing between brush and melt chains, and the loss of conformational entropy of the tethered chains, which stretch to allow the melt chains to come closer to the surface \cite{Leibler94, Macdowell05, Pastorino06}.  Both effects do not depend on the details of the interactions on the atomistic scale and can be described by coarse-grained polymer models.  Additionally, the two contributions to the free energy can be controlled in the simulations as well as in experiments by changing the molecular weight of the free molecules of the liquid or the tethered chains of the brush and the grafting density of the brush.

In the following we investigate a deceptively simple system comprised of polymer chains irreversibly tethered with one end to a solid substrate and a polymer melt of free polymer chains with the same molecular weight and interactions. In this way, the chains of the liquid and brush are of identical chemical nature.  The grafting substrate is perfectly flat and all friction between the substrate and the polymer melt stems from the interaction between the grafted chains of the brush and the liquid.

Varying the grafting density gives rise to a rich wetting behaviour \cite{Muller00f, Muller01b, Macdowell06, Pastorino06, Pastorino09}.  Three different regimes can be distinguished: (i) At very low grafting densities, wettability is controlled by the interactions between the solid grafting substrate and the liquid. If the solid substrate is sufficiently attractive, the liquid will wet the substrate and this wetting transition is of first order \cite{Muller01b}. Increasing the grafting density, the tethered chains provide additional attractive interactions for the liquid and the wetting transition occurs at a smaller value of the Hamaker constant, $A$, that characterise the strength of the long-range interaction between the solid substrate and the liquid. (ii) At intermediate grafting densities, the tethered chains provide enough attraction for the liquid to wet the brush even in the absence of additional attraction between the solid and the liquid. Long-range, second-order wetting transition can be observed as $A$ changes its sign \cite{Muller01b}. (iii) At high grafting densities, an interface between the brush and the melt gradually builds up and this interface becomes the narrower the larger the grafting density becomes. In this limit, the free chains of the melt cannot penetrate the brush and do not benefit from the attractive interactions between brush and melt. The brush-melt interface is characterised by a thermodynamic interface tension and the melt dewets from the brush although is comprised of identical constituents. This effect is denoted autophobicity \cite{Maas02, voronov03, voronov02}.

%
\begin{figure}[!t]
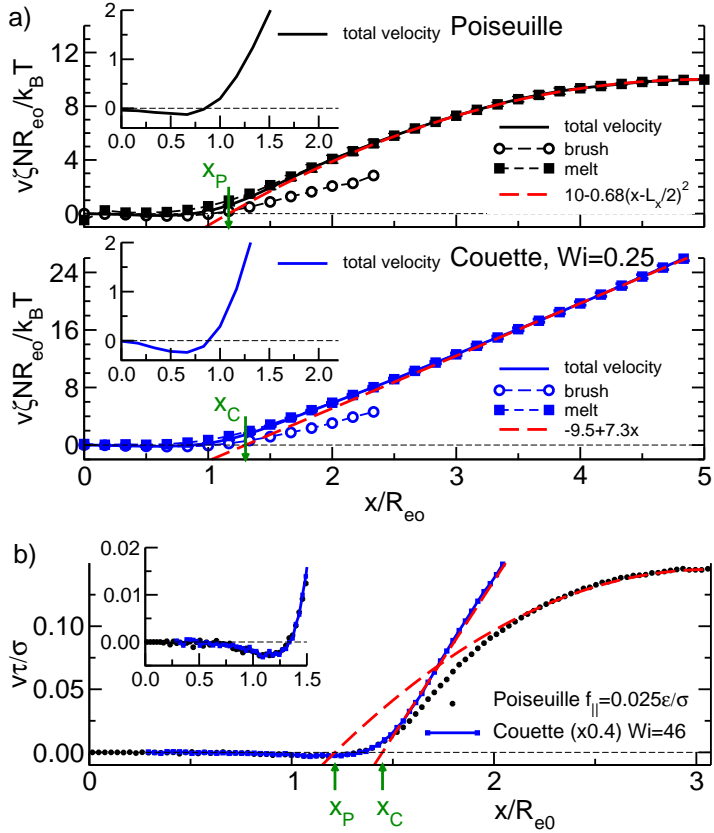

\includegraphics[clip,width=0.6\columnwidth]{fig5a.eps}\\
 \includegraphics[clip,width=0.6\columnwidth]{fig5b.eps}
\caption{ Flow of a polymer melt past a brush-costed surface. (a) Poiseuille and Couette flows using dynamic SCMF simulations. The solid line indicate the total velocity profile, dashed lines depicts the velocity of the brush and the non-grafted melt, respectively. (b) The same flows observed by molecular dynamics simulations of the coarse-grained model described in the text.  $x_{\rm P}$ and $x_{\rm C}$ indicate the position in the channel at which the fluid velocity is extrapolated to $0$. The insets highlight the region of the brush-melt interface in which inversion of the flow is observed for the total velocity profile. Adapted from \cite{Muller08}.}
\label{fig5} 
\end{figure}

The properties of the brush-melt interface have been studied by non-equilibrium molecular dynamics simulation of a Lennard-Jones, bead-spring model and dynamic SCMF simulations of a soft, coarse-grained polymer model \cite{Pastorino06, Pastorino09, Muller08}.  In figure \ref{fig5} we present the velocity profiles obtained from non-equilibrium simulations of the Lennard-Jones, bead-spring model for Couette flow and Poiseuille flow. First, we notice that the velocity profile of Couette flow extrapolates to zero at a position, $x_{{\rm C}}$, which is well inside of the channel, $x_{{\rm C}}>0$. Thus, one cannot simply identify the hydrodynamic position with the location of the solid grafting substrate, $x_{{\rm b}}=0$, because this identification would result in a negative value, $\delta=-x_{{\rm C}}<0$, according to \eqref{xb}. This negative value is inconsistent with the relation, $\delta=\eta/\lambda$, and the fact that both the shear viscosity, $\eta$, and the friction coefficient, $\lambda$ must be positive. More generally, this observation highlight that both parameters of the Navier-slip condition, $x_{{\rm b}}$ and $\delta$, have to be determined independently.

For a brush-coated substrate, we expect that for low grafting densities the hydrodynamic position, $x_{{\rm b}}$, is close to the solid substrate. At high grafting densities, however, the hydrodynamic position, $x_{{\rm b}}$, is dictated by the location of the interface between the brush and the melt. Quantitatively analysing the velocity profiles of Couette and Poiseuille flow, we again find that the parameters of the Navier-slip condition cannot be extracted because $x_{{\rm P}}>x_{{\rm C}}$ (cf.~figure \ref{fig5}) for a wide range of grafting densities \cite{Pastorino09}. This effect is observed for both, the Lennard-Jones, bead-spring model and the soft, coarse-grained model of the dynamic SCMF simulations, which correspond to quite different situations. Qualitatively, this failure of the Navier-slip condition can also be rationalised by the schematic, two-layer model: The grafted chains of the polymer brush dangle into the polymer liquid and increase the effective viscosity, $\eta_{{\rm S}}$, of the liquid that penetrates the brush. Milchev and co-workers have recently quantified the increase of the near-substrate viscosity by defining an effective local viscosity as the ratio between the local stress and the gradient of the velocity \cite{Dimitrov07, Dimitrov08c}. 

Alternatively, the motion of the fluid through the polymer brush has been conceived as the flow through a porous medium, which is described by Brinkmann's equation \cite{Milner91b, Brinkman47}. Describing the polymer brush as a static porous medium, which only imparts additional friction onto the melt, one would expect that the tethered chains tilt in the flow and slightly fluctuate around their average, tilted positions. In the simulations, however, we observe strong, non-Gaussian fluctuations of the molecular orientation in shear \cite{Muller08}. Similar, strongly non-Gaussian distributions of orientations have also been predicted for isolated, tethered chains under shear \cite{Winkler06}.

Only for extremely small grafting densities, $\sigma_{{\rm g}}$, we observe slippage because the solid substrate is assumed to be perfectly flat and in the limit $\sigma_{{\rm g}}\to0$ we observe plug flow with an infinite slip length, $\delta$ \cite{Pastorino06}. In the opposite limit of very high grafting densities, the interface between brush and melt becomes very narrow and the brush resembles a very dense solid with a rather smooth surface. In this extreme limit, we observe a finite slip length, $\delta$ \cite{Pastorino09}.

While the velocity profiles at the centre of a brush-coated channel exhibit the expected linear and parabolic velocity profiles for Couette and Poiseuille flow, respectively, we observe a reversal of the flow direction in the vicinity of the grafting surface. This small reversal is observed both in the dynamic SCMF simulations, which use rather small Weissenberg numbers, Wi, and large invariant degrees of polymerisation, $\Nbar$, as well as in simulations of the Lennard-Jones, bead-spring model, which utilise larger Wi and much smaller $\Nbar$ but capture the non-crossability of the flexible, linear chain molecules and the changes of the fluid structure in the vicinity of the solid substrate \cite{Muller08}.

The reversal of the flow direction is observed at intermediate grafting densities on the brush side of the well-developed brush melt interface.  In this region, the total velocity of the system is dominated by the velocity of the segments of the brush. Since the brush is irreversibly grafted to the solid substrate, which is at rest, the time average of the velocity of each brush segment in the steady state has to vanish. Nevertheless, the motion of the brush segment can couple to the rotational component of the shear flow at the brush-melt interface, which imparts a correlation between the tangential velocity of a grafted bead, $v_{\|}^{{\rm brush}}(x)$, and its distance, $x$, from the grafting surface. If a brush segment is located far away from the grafting surface, it is exposed to the flow and partially convected resulting in $v_{\|}^{{\rm brush}}(x)>0$ for large $x$. Since the time average of $v_{\|}^{{\rm brush}}$ vanishes, this must be compensated by a negative $v_{\|}^{{\rm brush}}(x)$ when the brush segment is located closer to the grafting substrate, \ie, small values of $x$.

%
\begin{figure}[!t]
\includegraphics[clip,width=0.6\columnwidth]{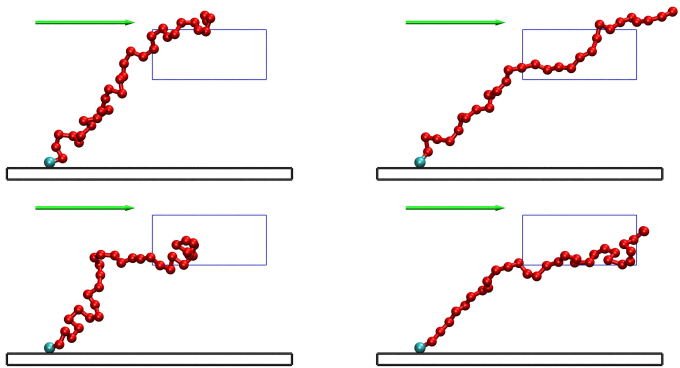}
\caption{$f_{\|}=0.04\epsilon/\sigma$ The chains are comprised of $N=32$ segments that interact via a repulsive Lennard-Jones potential with $r_{c}=2^{1/6}\sigma$. $k_{B}T/\epsilon=1.68$, $\sqrt{\Nbar}\approx6$ and $\sigma_{g}\R^{2}=1.5\Nbar$. The four panels depict a sequence of conformations extracted directly from a trajectory of the simulations.  (top-left) Thermal fluctuations expose the chain to regions of higher velocity and (top-right) the convection in the flow stretches the conformation parallel to the substrate. (bottom-right) When a fluctuation brings the chain closer to the substrate, it is exposed to a smaller or even reversed flow and (bottom-left) its lateral extension collapses. }
\label{fig6} 
\end{figure}

This cyclic motion of grafted polymers in shear flow has previously been predicted and observed for isolated, tethered molecules \cite{Doyle00,Gerashchenko06,Winkler06}. In our simulations we observe this cyclic tumbling motion in a polymer brush \cite{Muller08}. All chains perform this tumbling motion albeit in an unsynchronised manner. Their collective behaviour results in the reversal of the total flow direction in the vicinity of the grafting
substrate.

We expect that this inversion of the velocity is characteristic for intermediate grafting densities. At very small grafting densities, the flow at the surface is dominated by the velocity of the solvent.  The cyclic motion of the grafted chains will exist but it will not be sufficient to invert the total velocity, which is dominated by the flow of the solvent. Contrary, if the brush was strongly stretched, the width of the brush-melt interface would be narrow, the height fluctuations of brush segments would be small, and -- in the Lennard-Jones model and experiments -- topological constraints would become more important. These effects are expected to shift the inversion zone away from the grafting surface and tend to reduce the inversion of the flow.

\section{Polymer drops on surfaces}
The properties of the confining boundaries exert a pronounced influence on the flow behaviour at the ultimate vicinity of the surface. This is particularly true for nano- and micro-fluidic devices or small drops because of the larger surface-to-volume ratio. Boundary effects are difficult to predict in the framework of a continuum theory because particle properties, details of the interactions between particles and correlations become important (\eg, layer effects of a dense liquid at a surface). These local properties can be accurately and efficiently investigated by molecular dynamics simulation. 

In the section, we will explore, which parameters have to be determined from a microscopic, particle-based model, in order to describe the behaviour of drops on surfaces with slip. On the one hand, such a parameter-passing between molecular simulation and a continuum description is necessary in order to extrapolate the results of the molecular dynamics simulation to experimentally relevant length and time scales, which are difficult to assess by molecular dynamics simulations. On the other hand, the comparison between molecular simulations and continuum description using the independently determined parameters demonstrates that the continuum model captures the underlying physics.

\subsection{Static properties on hard and soft, deformable substrates}
First, we will consider the equilibrium properties of polymer droplets on a hard, corrugated substrate at high temperatures, $k_BT/\epsilon=1.2$ (cf.~section \ref{sec:CHS}). In this case, the hydrodynamic position agrees with the location of the interface between solid and liquid, and the slip length can be controlled by the attraction, $\epsilon_{\rm S}$, between solid and liquid. The macroscopic, equilibrium shape of a droplet is dictated by the volume, $V$ of the liquid and its contact angle, $\Theta$. The latter quantity can be estimated from the interface and surface tensions according to Young's equation.

Directly measuring the contact angle of microscopically small drops in the simulations, however, one faces multiple difficulties \cite{Muller03c}: (i) The liquid-vapour interface in the vicinity of the contact line is strongly distorted. In a cap-shaped drop geometry, these effects are particularly pronounced because the tension of the three-phase contact line between solid, liquid and vapour deforms the droplet. To leading order, the line tension, $\lambda$, gives rise to a dependence of the contact angle, $\Theta$, on the droplet radius, $R$, which takes the form of Gretz's equation \cite{Gretz66,Gretz66b}
\begin{equation}
\gamma_{\rm LV} \cos \Theta + \gamma_{\rm LS} - \gamma_{\rm VS} + \frac{\lambda}{R \sin \Theta} = 0
\end{equation}
(ii) The curvature of the liquid-vapour interface and the concomitant Laplace pressure shift the liquid-vapour equilibrium from its bulk coexistence value. Moreover, only in a certain range of system sizes equilibrium droplets of a specified radius, $R$, can be observed. \cite{MacDowell02b}. (iii) Microscopically, the contact angle is not well defined but it always involves an extrapolation of the position of the liquid-vapour interface towards the solid substrate. For small droplet sizes such an extrapolation imparts a significant error onto the estimates of the contact angle, $\Theta$.

\begin{figure}[!t]
\includegraphics[clip,width=0.6\columnwidth]{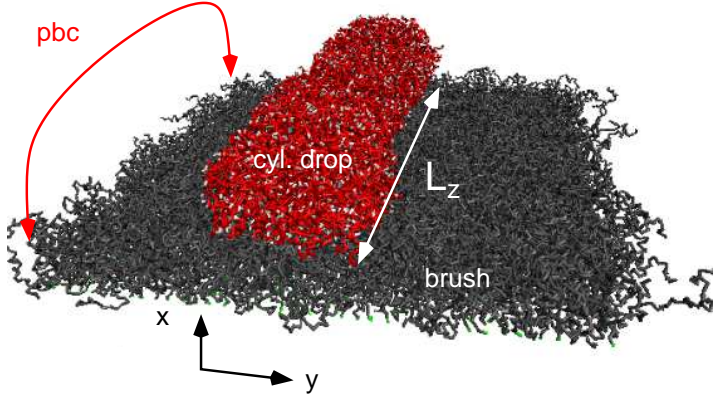}
\caption{Snapshot of a cylindrical polymer drop on a polymer brush. Brush segments are coloured in grey, grafted ends are depicted in light green. Segments of the melt are coloured in red. The figure corresponds to a snapshot of the simulation box for a grafting density $\rg\R^2=9.5$, chain length $N=40$, and $k_BT/\epsilon=1.2$ (with $r_c=2.5\sigma$).}
\label{fig7} 
\end{figure}

Therefore, we have decided to study cylindrical droplets as depicted in figure \ref{fig7}, which span the simulation cell in one direction. The length of the contact line is always $2L_z$ independent from the size of the droplet \cite{Neumanna1998, Kulmala2007, Servantie08}. This largely reduces the effect of the line tension, although minor effects due to the interaction of the liquid-vapour interface with the solid substrate remain. Moreover, the size of the cylindrical droplet, $R \sim n^{1/2}$ increases faster with the number of particles, $n$, than for a spherical drop, $R \sim n^{1/3}$. By virtue of the low temperature, vapour density and liquid compressibility are small, and the shift of the pressure away from the bulk coexistence value does not give rise to a noticeable change of the liquid or vapour properties. The third caveat can be mitigated by avoiding to extract the contact line by the details of the liquid-vapour interface at the three-phase contact but rather fitting the shape of the entire drop to the macroscopic prediction. For instance, one can measure the radius, $R$, of the drop and the height $x_{\rm CM}$, of its centre of mass as a function of the particle number, $n$. For a cylindrical drop, $R$ and $x_{\rm CM}$ are related to the contact angle, $\Theta$, via \cite{Servantie08}

\begin{equation}
\frac{x_{\rm CM}}{R} = \frac{4}{3} \frac{\sin^3 \Theta}{2\Theta-\sin 2\Theta}-\cos\Theta \label{STATRZ}
\end{equation}
Other quantities, like the volume, $V$, of the drop  and its contact area, $A$ can be expressed as

\begin{eqnarray}
\frac{V}{R^2L_z} &=& \frac{2\Theta-\sin 2 \Theta}{2} \label{STATVOL}\\
\frac{A}{RL_z} &=& 2 \sin \Theta \label{STATAREA}
\end{eqnarray}
Fitting those relationships as a function of $n$, one can rather accurately determine the contact angle, $\Theta$, and the droplet radius, $R$, from molecular simulations of microscopically small, cylindrical droplets. The translational symmetry along the $z$-axis allows to average profiles and increases the statistics of the simulation data.

%

If the substrate is soft, then the forces that act at the three-phase contact line will also deform the substrate and lift up the three-phase contact line. This effect is well-known for liquid substrates and it is also notable for soft solids, like brush-coated substrates or polymer networks. The free energy penalty of a surface deformation of molten brushes has been considered within the strong stretching approximation. These calculations typically consider an incompressible brush and a sharp brush-melt interface. Under these conditions height fluctuations of the brush with long lateral wavelengths will be strongly suppressed because of the prohibitive costs of stretching of the lateral chain conformations \cite{Fredrickson92b,Xi96,Likhtman99}. In our simulations, however, the polymer liquid exhibits a non-trivial equation of state with a finite compressibility. Therefore, a deformation of the surface of the brush can result in density fluctuations and even a spatially homogeneous displacement of the brush-melt interface is possible \cite{Pastorino09}.

The deformability of the brush surface decreases with the grafting density, $\rg$, because (i) the number of chains, which are stretched, increases and (ii) the density inside the brush increases and its compressibility decreases in turn. We have tried to study the deformability at the three-phase contact line of a polymer brush and a droplet that is comprised of identical chains. In order to achieve a contact angle of about $90^o$, the grafting density has to be large. Under these conditions the brush is hardly deformable and resembles a solid substrate. If the grafting density is low, in turn, the brush will be deformable but the polymer liquid will wet the brush or will make a very small contact angle. In this case, the forces at the three-phase contact line are nearly tangential and the lifting-up of the contact line is minuscule.

%
\begin{figure}[!t]
\includegraphics[clip,width=0.6\columnwidth]{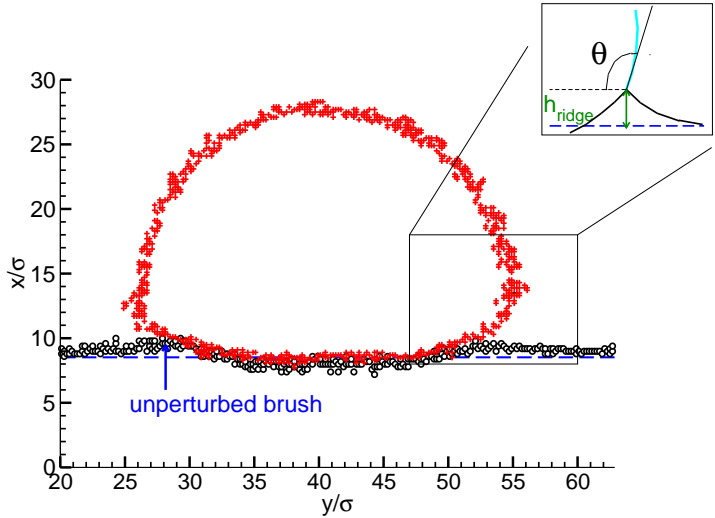}
\caption{Density contour plot of a $N=10$ mers polymer liquid droplet wetting a deformable substrate. The reduced grafting density of the polymer brush substrate is $\rg \R^2=13.1$ and the brush-melt compatibility is $\epsilon_{\rm bm}=0.6\epsilon$. The dashed line represents the average free surface position of the unperturbed brush. The lifting-up of the contact line and the formation of a ridge is visible.}
\label{fig8} 
\end{figure}

Therefore, we have decided to make the molecules of the brush and the polymer liquid slightly incompatible by reducing the energy parameter of the Lennard-Jones potential between segments of the brush and melt, $0.3 \epsilon \leq \epsilon_{\rm bm} \leq 0.8 \epsilon$. Varying $\epsilon_{\rm bm}$ and the grafting density, $\rg$, we can independently control the contact angle, $\Theta$, and the deformability of the polymer brush, and thereby identify conditions, where the brush is soft and the liquid-vapour interface is perpendicular to the substrate at the three-phase contact line. Figure \ref{fig8} depicts the density contours of such an incompatible polymer drop comprised of chains with $N=10$ on a slightly incompatible polymer brush. 

%
\begin{figure}[!t]
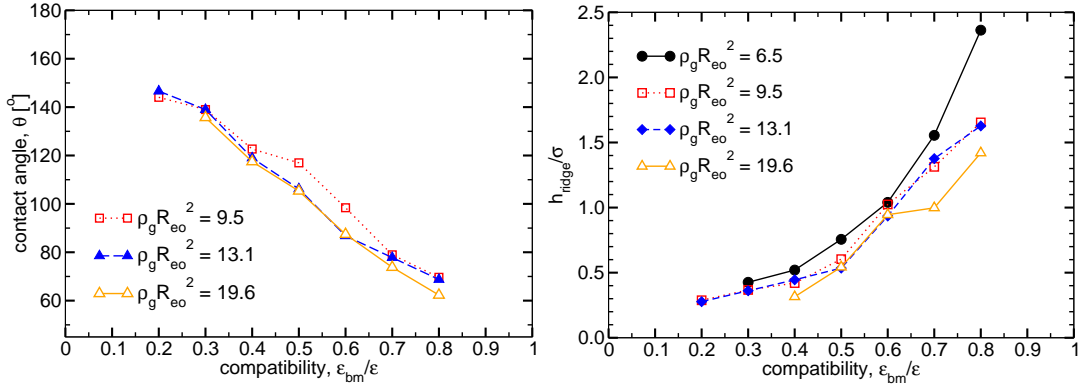

\includegraphics[clip,width=0.45\columnwidth]{theta_ellip.eps}
\includegraphics[clip,width=0.45\columnwidth]{h_lift_up_rescaledRe.eps}
\caption{Dependence of the shape of a polymer drop ($N=10$) on a brush ($N=40$) as a function of their compatibility, $\epsilon_{\rm bm}$. The left panel (a) presents the contact angle, $\Theta$, for different grafting densities, while the right panel (b) depicts the height, $h_{\rm ridge}$ of the ridge at three-phase contact.}
\label{fig9} 
\end{figure}

From these contour plots one can roughly estimate the contact angle, $\Theta$, of the polymer drop on top of the brush. These data are presented in figure \ref{fig9}a for a brush composed of chains of lengths, $N_{\rm b}=40$, and a melt of short polymers, $N=10$. As one increases the compatibility between the brush and the melt, $\epsilon_{\rm bm} \to \epsilon$, the contact angle decreases. For the value,  $\epsilon_{\rm bm} \approx 0.6 \epsilon$, we achieve a contact angle of $90^o$. Interestingly, the data show that the contact angle is rather independent from the grafting density of the brush. We also find that $\Theta$ does not significantly vary with the length of the grafted chains (data not shown). For a brush with an intermediate grafting densities in a bad solvent, the structure of the brush surface is independent from $\rg$ and resembles a liquid-vapour interface. The independence of the contact angle therefore implies that the brush-melt interface does not significantly change with grafting density, $\rg$. Thus, it appears that the value of $\epsilon_{\rm bm} \approx 0.6 \epsilon$ is sufficient to dominate the tension between the brush and the melt and to outweigh the subtle  entropic contributions that determine autophobicity and that do depend on grafting density and the ratio of chain lengths between the grafted and free chains \cite{Leibler94}.

The height, $h_{\rm ridge}$, of the ridge as a function of the compatibility, $\epsilon_{\rm bm}$, is presented in panel (b) of figure \ref{fig9}. It is only of the size of the segment diameter, $\sigma$, and it slightly increases with reducing the grafting density because the brush becomes softer. The main deformation, however, occurs at the brush-melt interface and thus the dependence on $\rg$ again is rather weak. Although the contact angle decreases with $\epsilon_{\rm bm}$, and therefore the tension of the liquid-vapour interface acts more tangentially to the surface, the height, $h_{\rm ridge}$, of the ridge increases as we make brush and melt more compatible. Note also that on the microscopic scale of the range of interaction, $r_c = 2.5 \sigma$, (cf.~figure \ref{fig8}), the liquid-vapour interface bents around before the ridge, where it meets the substrate. 

\subsection{Motion of droplets on hard substrates}

To illustrate that the microscopic details of the structure at the liquid solid interface can be accounted for in a hydrodynamic description by the contact angle and the slip length, we have studied the motion of droplets under the influence of an external body force \cite{Servantie08}. Examples of such a situation are falling drops on an inclined plane due to gravity \cite{Pomeau1999, Pomeau2001, Thiele02, Mognetti10}. Much attention has been devoted to the onset of motion, the changes of the shape of the droplets, and their instabilities in response to strong body forces \cite{Huppert82, Thiele02, Rio05, LeGrand05, LeGrand06b}. Comparably less attention has been devoted to the steady-state rolling and sliding of drops in external fields \cite{Kim02d, Mognetti10}, which is pertinent to nano-fluidic devices. 

The steady-state tells a great deal about the dissipation mechanisms inside the moving droplets for the drop's velocity is dictated by the balance between the power gained through the motion in the external field and the dissipation. The dominating source of dissipation depends on the size of the droplet \cite{BrochardF1992} and a systematic study of the size dependence is indispensable for identifying the different contributions in a simulation study. In the following we use the same model as in section \ref{sec:CHS} on two layers of an FCC Lennard-Jones solid.

Similar dissipation mechanisms also act when a droplet spreads on a wettable substrate \cite{Tanner79, Ruijter97, Ruijter99, Milchev02e, Heine03, Heine04, Webb2005, Heine05b} but the study of the steady-state properties is computationally much more convenient because one can average properties along the trajectory instead of averaging over many realisations of the spreading process.

The total dissipation rate, $T\sum_{\rm TOT}$ is additively comprised of different contributions \cite{deGennes1985}. Energy is dissipated in the flow of the viscous fluid inside the drop. Assuming that the velocity field is predominantly tangential to the substrate, the lubrication approximation estimates

\begin{equation}
T\sum_w \approx \eta \int_V {\rm d}V \; \left(\frac{\partial v_\|}{\partial x}\right)^2 \approx \xi \eta \left(\frac{v_{\|{\rm S}}}{\delta}\right)^2 V
\label{DISSEQ6}
\end{equation}
This contribution scales like the volume, $V$, of the drop. In the last step, we have rather crudely estimated the typical velocity gradient, $\frac{\partial v_\|}{\partial x}$ inside the drop by its value, $\frac{v_{\|{\rm S}}}{\delta}$, at the substrate according to the Navier-slip condition \eqref{Navier}. $\xi$ is a factor of order unity that describes the details of the velocity field inside the drop. The latter approximation is restricted to small contact angles and large slip lengths, $\delta$.

An additional source of dissipation are processes at the three-phase contact line \cite{Blake69, Bertrand07, Voue07}, which {\em inter alia} may arise from pinning at the contact line or evaporation/condensation of liquid at the three-phase contact. This contribution, however, scales only like the length of the three-phase contact line. In the case of cylindrical droplets it provides a constant dissipation independent of the drop size and even for cap-shaped drops we expect this contribution to only influence the behaviour of the smallest droplets. If the substrate was soft and deformable (\eg, a brush-coated solid), then the lifting-up of the three-phase contact and the deformation of the soft solid in the course of the motion would constitute an additionally dissipation mechanism at the contact line \cite{Shanahan88, Carre93, Shanahan94, Shanahan95, Carre96}. This effect has been termed ``viscoelastic braking''\cite{Carre93} and is an interesting subject for future simulation studies using the substrates characterised in the previous subsections.

Another contribution may arise from a precursor film that spreads in front of the drop. Dissipation in such a precursor film is important for spreading of drops but it is not observed in the present simulation of the motion of drops under the influence of a body force. Therefore it is not considered further in the following.

If slip occurs at the solid-liquid boundary, however, an additional important dissipation mechanism arises that stems from the friction as the liquid slips past the solid. Since the entire contact area, $A$, moves, this dissipation mechanism scales like $A$ and outweighs the dissipation at the three-phase contact. In the ultimate vicinity of the corrugated substrate, on the microscopic scale, the liquid flow follows the substrate shape and the concomitant velocity gradients result in dissipation. On the one hand, this friction can be conceived as a viscous dissipation at the solid-liquid boundary. On the other hand, the physical mechanisms of friction at the  solid-liquid boundary resemble the processes at the three-phase contact line.\cite{Blake69} Although the origin of friction at the solid-liquid boundary is complex and possibly diverse, its effect can be quantitatively described via the slip length, $\delta$. The power dissipated by the friction is
\begin{equation}
T\sum_A=F_{\rm fric} v_{\|{\rm S}} \qquad \mbox{with} \quad F_{\rm fric}=\eta \frac{v_{\|{\rm S}}}{\delta} A
\label{DISSEQ4}
\end{equation}
where the dynamic friction force, $F_{\rm fric}$, is given by the definition of the Navier-slip \eqref{Navier} \cite{Bocquet1994} and $v_{\|{\rm S}}$ is the velocity of the liquid at the substrate.

%
\begin{figure}[!t]
\includegraphics[clip,width=0.45\columnwidth]{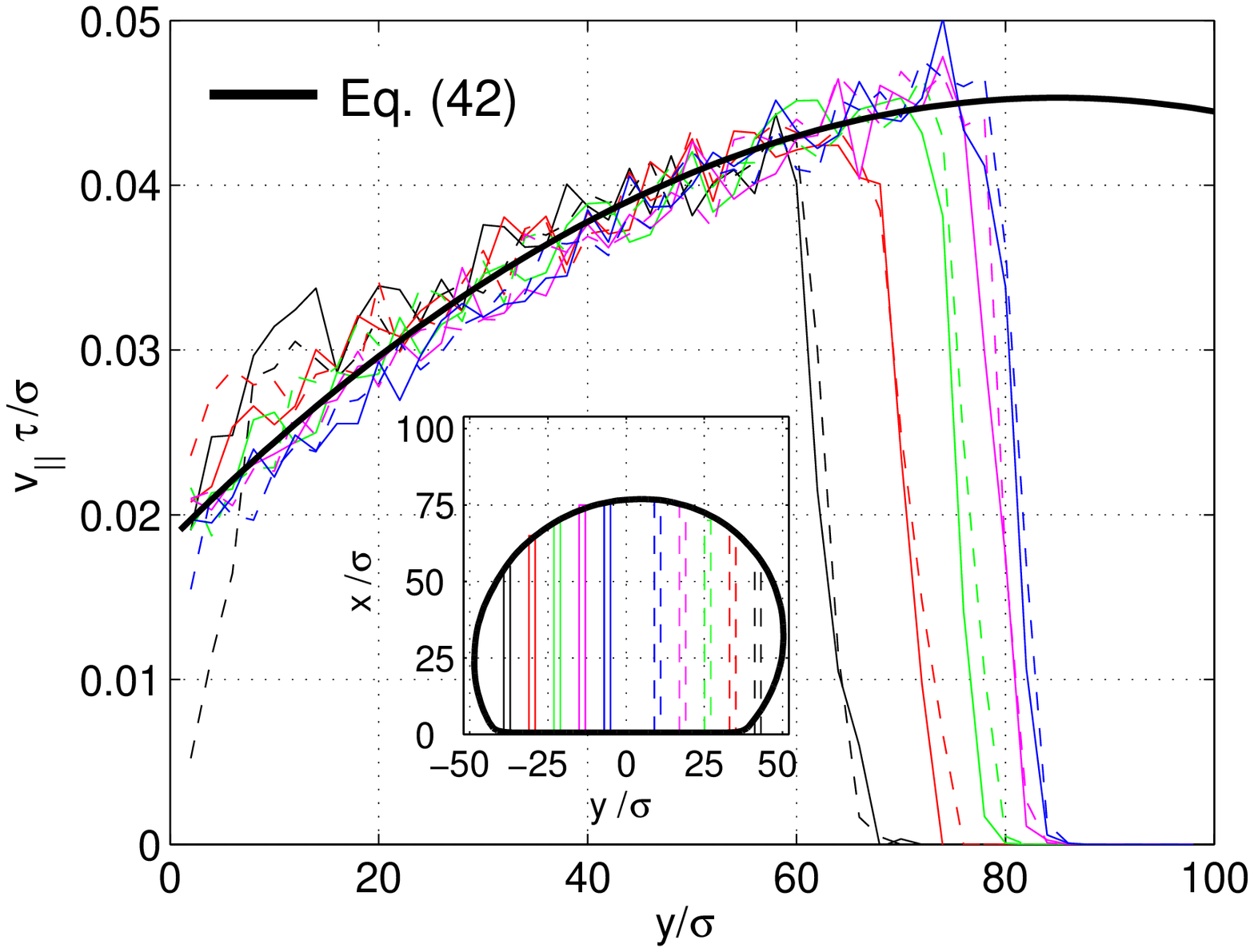}
\includegraphics[clip,width=0.51\columnwidth]{fig10.eps}
\caption{
a) Velocity profile as a function of the height, $x$ for different lateral positions, $y$, inside the droplet. The inset shows the sections, which have been
used to calculate the velocity profile $v_\|(x,y)$.
b) Velocity of the centre of mass per unit acceleration as a function of the number of monomers for $\epsilon_{\rm S}=0.4\ \epsilon$ ($\Theta = 130^o$, upper panel) and $\epsilon_{\rm S}=0.8\ \epsilon$ ($\Theta \approx 29^o$, lower panel). The circles are the results of the molecular dynamics simulations and the dashed lines correspond to \eqref{vcem} using the independently determined values of $\eta$ and $\delta$. The error bars mark the uncertainties associated with these independently determined input parameters. The asymptotic limits for small and large droplets are indicated by the solid lines proportional to $\sqrt{n}$ and $n$, respectively.  
Adapted from Ref.~\cite{Servantie08}
}
\label{fig10}
\end{figure}

From dimensional considerations, \eqref{DISSEQ6} and \eqref{DISSEQ4} imply that the dissipation in large drops is dominated by the viscous dissipation of the liquid flow inside the drop while for smaller droplets that friction at the substrate plays in important role. Since the slip length, $\delta$, is the only length scale the ratio, $R/\delta$, separates large drops from small droplets. Using the approximation for the viscous dissipation in the volume of the drop, one can make further progress by balancing the power gained by the motion of the drop with velocity, $u$, in the external body force, $\rho_{\rm L} f_\|$, against the two dominating dissipation mechanisms

\begin{equation}
\rho_{\rm L} V f_\| u = \eta \frac{v_{\|{\rm S}}^2}{\delta} A + \xi \eta \left(\frac{v_{\|{\rm S}}}{\delta}\right)^2 V
		      \sim \eta \left(\frac{u}{\delta + x_{\rm CM}}\right)^2 ( \delta A + \xi V)
\end{equation}
where, in the last step, we have estimated the magnitude of the velocity of the drop, $u$, by the tangential velocity of the liquid inside the droplet at the height, $x_{\rm CM}$, of its centre of mass, $u \approx v_\|(x_{\rm CM}) \approx v_{\|{\rm S}} + \frac{v_{\|{\rm S}}}{\delta} x_{\rm CM}$. This relation implies that the steady-state velocity, $u$ is given by
\footnote{
The velocity, $u$, and the body force, $\rho_{\rm L} f_\|$, can be quantified by the dimensionless capillary number, Ca$\equiv u\eta/\gamma_{\rm LV}$, and the Bond number, Bo$\equiv \rho_{\rm L}f_\|R^2/\gamma_{\rm LV}$. Then \eqref{vv} can be rewritten in the form $\eta u/(\rho_{\rm L}f_\|R^2) \sim$ Ca$/$Bo $\sim \delta/R$ for small droplets while the ratio Ca$/$Bo approaches a constant for large drops.
}
\begin{equation}
\frac{\eta u}{\rho_L f_\|} \sim \left\{
				\begin{array}{lll}
				\delta R    & \mbox{for}\, \delta A \gg V  & \mbox{dominated by friction}           \\
				R^2  & \mbox{for}\, \delta A \ll V  & \mbox{dominated by viscous dissipation}
				\end{array}
                                \right. 
\label{vv}
\end{equation}
A somewhat more quantitative description of the velocity field inside the drop can be obtained by assuming that the tangential velocity profile, $v_\|(x)$,  only depends on the distance, $x$, from the substrate. In this case, the Stokes equation for a liquid under the influence of a body force predicts a parabolic velocity profile

\begin{equation}
v_\|(x)=\frac{\rho_{\rm L} f_\|}{\eta}\left[\left(H-\frac{\Delta x}{2}\right)\Delta x+\delta H\right]
\label{vcem}
\end{equation}
where $\Delta x = x - x_{\rm b}$ denotes the distance from the hydrodynamic boundary position, and $H$ is the height of the drop with respect to $x_{\rm b}$. The coefficient in front of the quadratic term is dictated by the driving force, $f_\|$, and the remaining two coefficients have been obtained from the Navier-slip condition \eqref{Navier} and the observation that there is no tangential stress at the liquid-vapour interface, \ie, $\frac{\partial v_\|}{\partial x}=0$ for $\Delta x = H$. In figure \ref{fig10} we compare this phenomenological hydrodynamic description with our simulation data. In panel (a) we demonstrate that the tangential velocity profile, $v_\|(x)$, is indeed largely independent from $y$ and that \eqref{vcem} provides a rather accurate description using the previously determined parameters, $\eta$ and $\delta$. Panel (b) depicts the dependence of the drops velocity, $u \approx v_\|(x_{\rm CM})$, on the size of the cylindrical drop and compares the simulation results to \eqref{vcem} without adjustable parameter. Quantitative agreement is found. The description also captures the counter-intuitive observation that the crossover from the friction-dominated behaviour of small droplets to the viscosity-dominated behaviour of large drop occurs at smaller sizes, $R_c \sim \delta/(1-\cos \Theta)$ for the larger contact angles, \ie, the crossover between surface-dominated and volume-dominated dissipation occurs at a larger volume, $n_c \approx 160\,000$, for systems with a larger contact angle ($\epsilon_{\rm S} = 0.4 \epsilon$) than for flatter droplets ( $n_c \approx 33\,000$, $\epsilon_{\rm S} = 0.8 \epsilon$). Although, at fixed volume, the flatter drop makes a larger area with the solid substrate, the velocity field inside a flat drop also strongly deviates from that of a solid-body rotation and therefore significantly contributes to the dissipation.

\subsection{Outlook on dewetting of polymer films}

%
\begin{figure}[!t]
\includegraphics[clip,width=0.9\columnwidth]{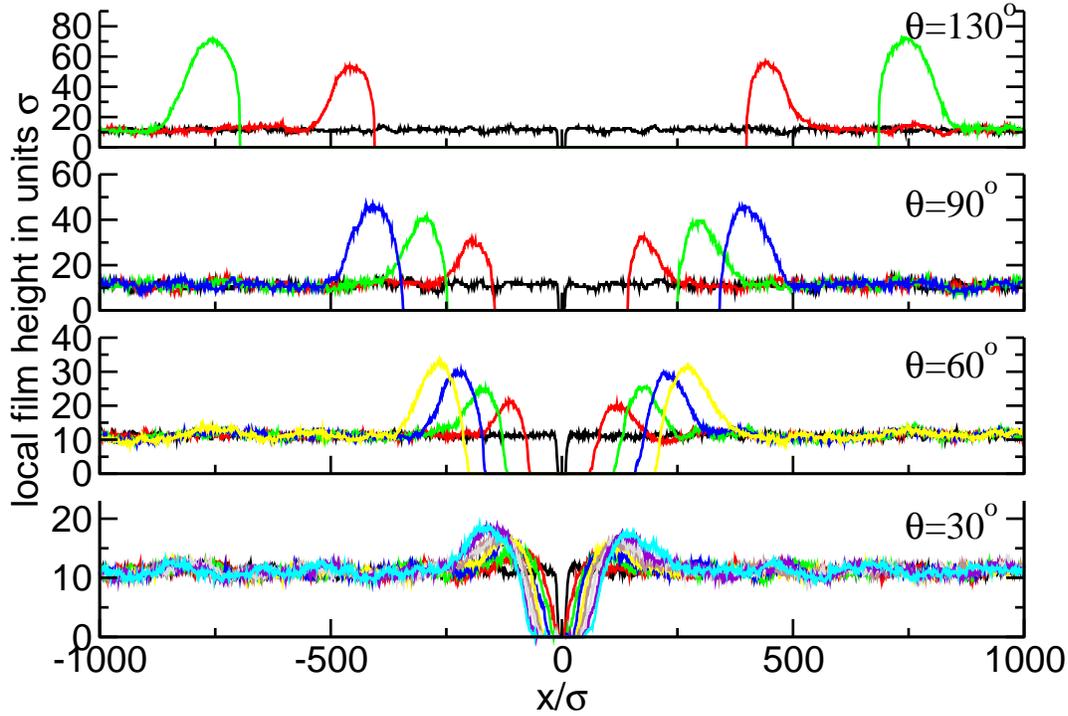}
\caption{Dewetting profiles for different equilibrium contact angles as indicated in the key. The length of the substrate is $2000 \sigma$ and its width $20 \sigma$. Each system is comprised of $300\,000$ particles. The successive profiles correspond to snapshots at $5000\;\tau$ time intervals.}
\label{fig_dewet}
\end{figure}

In equilibrium a partially wetting liquid on a solid substrate has the shape of a drop with a contact angle, $\Theta$, determined by Young's equation, \eqref{Young}. However, the polymer liquid can be initially prepared as a metastable film of uniform thickness on the substrate (\eg, by spin-coating). In this case, two situations can be distinguished: (i) If the film is thin enough, the fluctuations of the local film thickness in the film will spontaneously grow everywhere in the film. Through this spinodal dewetting mechanism \cite{Xie98}, the film will rapidly relax to its equilibrium configuration (\ie, a drop). (ii) A thick film will be metastable to fluctuations, and the dewetting process can be initiated on relevant time scale only by defects or a rupture of the film. The simulation data in figure \ref{fig_dewet} illustrate this second dewetting scenario. In this case a hole is nucleated and its radius grows in time. 

Both, the shape of the rim that forms as the liquid retracts from the hole and the dewetting speed, by which the hole's radius grows, depend on the microscopic properties of the substrate \cite{Brochardwyart94b, Fetzer06, Fetzer07c, Fetzer07d}, the initial film thickness \cite{Reiter92, Reiter93, Muller01, seemann01b} and the viscous properties of the liquid \cite{Reiter01b, Herminghaus02, Rauscher05, Munch06, Blossey06, Baumchen2009}. The larger the equilibrium contact angle the faster the dewetting, and this effect is also observed in the data depicted in figure \ref{fig_dewet}. Moreover, slippage of the liquid past the substrate determines the dewetting speed and rim profile \cite{Blossey06, Fetzer05, Fetzer06, Fetzer07c, Fetzer07d}. It can be shown that for plug flow the hole radius, $R$, increases as $R \sim t^{2/3}$, while for a stick boundary condition one finds $R \sim t$ \cite{Brochardwyart94b}. As for droplets, this will cause a transition from surface-dominated to volume-dominated regime during the growth of the rim. The presence or absence of slip also affects the shape of the rim; indeed the rim profile can decay towards the outer, unperturbed region with oscillatory patterns as observed in experiments and predicted by analytical models \cite{Fetzer07c}. Thus, a great deal of information about slippage and the hydrodynamic boundary conditions can be extracted from the speed and the shape of the dewetting front observed in experiments.

Molecular dynamics simulation permits to have a microscopic insight into the dewetting dynamics and compare to the predictions of analytical models using the independently determined parameters like viscosity, surface tension and slip length, Thus one can gauge the validity of the necessary approximations in the analytical description of the macroscopic behaviour and indicate oversimplification (\eg, neglect of thermal fluctuations).

Unfortunately the simulation of dewetting phenomena poses a huge computational challenge because they require the study of large systems with more than $10^5$ beads over long times in order to observe the transition from slip-dominated to viscosity-dominated dewetting behaviour. Using efficient simulation techniques such studies have just become feasible as illustrated in figure \ref{fig_dewet}. Already the preparation of the initial starting configuration of a homogeneous, thick polymer film is computationally difficult. Knowing the single-chain statistics and by using optimised packing algorithms, one can overcome this obstacle \cite{Homberg08} and prepare a thin, equilibrated polymer film that is comprised of $300\,000$ segments on a hard solid substrate with FCC structure. Periodic boundaries in the $y$ and $z$ directions of dimensions $2000\;\sigma$ and $20\;\sigma$ are applied and the unperturbed film thickness amount to $h=12\;\sigma$. By varying the strength of the substrate $\epsilon_{\rm S}/\epsilon$ between $0.4-0.8$ one controls the equilibrium contact angle and slip length, and consequently the dewetting speed. Once the film equilibrated, we remove polymers at the centre in a slab of length $10\;\sigma$ and start the simulation. After a fast relaxation at the three-phase contact line the film starts to dewet the substrate. We depict in figure \ref{fig_dewet} the successive steps of dewetting from which the time dependence of the hole growth and the shape of the rim can be analysed similar to the experimental studies \cite{Fetzer07c}. This analysis is ongoing.

One important difference between the experiment and the simulations, however, is already visible in figure \ref{fig_dewet}. In the experimental system, the profiles of the rim of the hole are perfectly smooth and fine details of the rim shape can be analysed from a single snapshot. In contract to the spinodal dewetting process \cite{Fetzer07b}, thermal fluctuations are not of crucial importance once a hole has formed. The length scale in the simulation, however, is much smaller and on these microscopic length scales thermal fluctuations do matter. For a single snapshot, these thermal fluctuations of the liquid-vapour interface, \ie capillary waves, are difficult to distinguish from the shape of the rim predicted by analytical approaches without fluctuations. In particular, one would have to average the simulations over multiple realisations of the dewetting process in order to decide if the profile of the rim decays oscillatory towards the initial film thickness.

\section{Perspectives}
Molecular simulations can build a bridge between the microscopic flow behaviour of particle-based models and a continuum description. While those multi-scale techniques have matured for equilibrium properties \cite{Muller-Plathe02, IZVEKOV05_2, Praprotnik05, Praprotnik08, Lu10}, the development of computational strategies that can describe the dynamics of polymer liquids across different scales is still in its infancy. On the one hand, no single computational technique can simultaneously address the length and time scales from the microscopic structure and dynamics at the boundary between the solid substrate and the liquid to the motion of liquid drops in nano- or micro-fluidic devices. On the other hand, the microscopic details of the interactions dictate the behaviour on the larger scales. For instance, the microscopic interactions between atoms determine the viscosity of a liquid. This shear viscosity, in turn, controls how much liquid will pass through a pipe or a microfluidic channel in response to a pressure drop. In order to achieve this goal, one can use (i) a coherent multi-scale simulation scheme \cite{Delgado-Buscalioni03b, Fabritis06, Koumoutsakos05, Delgado-Buscalioni08}, which explicitly couples the particle-based microscopic model and a hydrodynamic, continuum description or (ii) a parameter-passing technique. We have adopted the latter approach in this work. 

The underlying idea is that the complex, microscopic behaviour inside the liquid and at its boundaries can be captured by a small number of coarse-grained parameters that, in turn, enter the hydrodynamic description. These coarse-grained parameters depend in a complex manner on the microscopic interactions, and it is the basic {\em assumption} that the behaviour on large time and length scales depends on the microscopic properties only {\em via} these coarse-grained parameters. The bulk behaviour of a simple, one-component liquid is characterised by its density, $\rho_{\rm L}$, and its shear viscosity, $\eta$. In order to describe the behaviour at its boundaries, we have used the contact angle, $\Theta$, (or the free energy cost of a unit area of surface) as well as the slip length, $\delta$, and the position, $x_{\rm b}$ of the hydrodynamic boundary. Young's equation, \eqref{Young}, and the Navier-slip condition, \eqref{Navier}, and the fact that these parameters enter into the continuum description of liquids in terms of the Navier-Stokes equation or approximations thereof (\eg, lubrication approximation \cite{Oron97}) have motivated this specific choice of parameters.

We would like to emphasise that this specific choice of coarse-grained parameters, $\rho_{\rm L}$, $\Theta$, $\eta$, $\delta$, and $x_{\rm b}$, that are necessary to transfer knowledge from the microscopic model onto the level of a continuum description is not unique. This list depends on the specific problem that one wants to describe. For instance, for strong shear, the non-Newtonian character of polymer liquids will become important, which may be captured by a dependence of the viscosity on the shear rate (shear thinning) or frequency (visco-elasticity). Under strong non-equilibrium conditions the temperature may not be constant in the liquid and the heat flux has to be considered. In the simplest case this will add the thermal conductivity to the list of coarse-grained parameters. There may be also additional parameters required to capture the behaviour at the boundary. In order to describe the details of the liquid-vapour interface in the vicinity of the contact line, \eg, one needs the interface potential, $g(h)$, which measures the free energy cost per unit area of placing the liquid-vapour interface a distance, $h$, away from the solid substrate. 

The role of molecular simulations in this scope is two-fold: (i) Measurement of these coarse-grained parameters for a specific microscopic model: Such studies are practically useful because they predict the behaviour of liquids. An understanding of the molecular basis of the coarse-grained parameters will contribute to tailoring the properties of boundaries between solid and liquids. This aim requires accurate and efficient computational techniques for calculating the bulk and boundary properties for molecular model of liquids via simulations. For instance, several simulation techniques for calculating the shear viscosity (\eg, Green-Kubo relation, \eqref{eq:visco_GK}, and reverse non-equilibrium molecular simulations, \eqref{FMP}, \cite{MullerPlathe99}) and extracting the parameters of the hydrodynamic boundary condition have been discussed. (ii) Testing the validity of the parameter-passing strategy: By a judicious selection of model systems that incorporate the relevant aspects of the physical behaviour on the molecular scale, one can quantitatively compare molecular simulations and continuum description on scales where both descriptions are valid. First, these studies provide information down to which length scale a continuum description is valid and how it breaks down. Second, they allow to explore to what extent the choice of coarse-grained parameters will be sufficient to capture the relevant physics \ie, ask the question: What are the relevant coarse-grained parameters that transfer the physical properties at the molecular level into the continuum model? This question can be only answered for specific systems. 

For the motion of liquid drops with a rather large slip length on a hard, corrugated substrate, our simulations suggest that the coarse-grained parameters, which we have listed above, yield an appropriate description of the microscopic physics because using the independently determined parameters in an approximate hydrodynamic description provides a quantitative description of the velocity profile inside the drop and its steady-state velocity \cite{Servantie08}. A detailed comparison with a numerically accurate solution of the hydrodynamic theory for the steady-state motion is certainly warranted to validate this result and to help extrapolating the results to larger, experimentally accessible drop sizes.

In the example of a brush-coated surface \cite{Muller08, Pastorino09} or a strongly attractive solid \cite{Servantie08b}, our simulations suggest that the slip length, $\delta$, and hydrodynamic position, $x_{\rm b}$, do not provide an appropriate description of the flow at the boundary. The simulations demonstrate that these two parameters are not able to simultaneously describe planar shear flow (Couette) and parabolic, pressure-driven flow (Poiseuille). This failure of the Navier slip condition can be ``reproduced'' by a schematic, two-layer model demonstrating that it is not related to a shear-rate dependence of the viscosity (shear thinning) or a time-dependent viscosity (visco-elasticity). We speculate that it truly is a boundary effect and not related to bulk properties of the fluid (\eg, non-Newtonian character of the polymer liquid). It is rather tempting to attribute this failure to higher-order spatial derivatives in the relation between viscous stress at the substrate and stress at the boundary.

Since the question about the appropriate coarse-grained parameters for fluid flow at boundaries can only be answered for specific systems there is a rich and wide-open spectrum of systems and processes to be investigated. In this manuscript we have restricted our attention to the most basic, coarse-grained models of polymer liquids (\ie, a Lennard-Jones bead-spring model and a bead-spring model with soft potentials) and the most basic flow phenomena (Couette and Poiseuille flow, and the steady-state motion of droplets in response to a body force). Of course, nano- and microfluidic systems offer a much greater variety of  flow geometries (\eg, chemically or topologically structured surfaces), deal with more complicated fluids (\eg, viscoelastic liquids), and often require to consider additional phenomena (\eg, spreading of drops or mixing in multi-component liquids, evaporation effects). It is an open, challenging, and very important question to generalise the concept of parameter-passing to these more complex situations.

\subsection*{Acknowledgments}

It is a great pleasure to thank M.P.~Allen, K.~Binder, K.Ch.~Daoulas, M.~H{\"o}mberg, K.~Jacobs, T.~Kreer, L.G.~MacDowell, M.~M{\"u}layim, P.~M{\"u}ller-Buschbaum, A.~M{\"u}nch, R.~Seemann, N.~Tretyakov, and B.~Wagner for fruitful collaborations and stimulating discussions. This work has been supported by the priority program ``Nano- and microfluidics'' of the German Science Foundation (DFG) under grant Mu1674/3. Ample computer time at the computing centres JSC J{\"u}lich, HLRN Hannover and Berlin, as well as the GWDG G{\"o}ttingen are gratefully acknowledged.

\section*{References}

\begin{thebibliography}{100}

\bibitem{Young}
T.~Young.
\newblock An essay on the cohesion of fluids.
\newblock {\em Philos.\ Trans.\ R.\ Soc.\ London}, 5:65--87, 1805.

\bibitem{Cahn77}
J.~W. Cahn.
\newblock Critical-point wetting.
\newblock {\em J. Chem. Phys.}, 66:3667--3672, 1977.

\bibitem{deGennes1985}
P.~G. de~Gennes.
\newblock Wetting: statics and dynamics.
\newblock {\em Rev. Mod. Phys.}, 57:827, 1985.

\bibitem{Sikkenk87}
J.~H. Sikkenk, J.~O. Indekeu, J.~M.~J. Vanleeuwen, and E.~O. Vossnack.
\newblock Molecular-dynamics simulation of wetting and drying at solid-fluid
  interfaces.
\newblock {\em Phys. Rev. Lett.}, 59:98--101, 1987.

\bibitem{Vanswol91}
F.~van Swol and J.~R. Henderson.
\newblock Wetting and drying transitions at a fluid-wall interface -
  density-functional theory versus computer-simulation .2.
\newblock {\em Phys. Rev. A}, 43:2932--2942, 1991.

\bibitem{Adams91}
P.~Adams and J.~R. Henderson.
\newblock Molecular-dynamics simulations of wetting and drying in {LJ} models
  of solid fluid interfaces in the presence of liquid vapor coexistence.
\newblock {\em Mol. Phys.}, 73:1383--1399, 1991.

\bibitem{Ellis04}
J.~S. Ellis and M.~Thompson.
\newblock Slip and coupling phenomena at the liquid-solid interface.
\newblock {\em Phys. Chem. Chem. Phys.}, 6:4928--4938, 2004.

\bibitem{Quake2005}
T.~M. Squires and S.~R. Quake.
\newblock Microfluidics: Fluid physics at the nanoliter scale.
\newblock {\em Rev. Mod. Phys.}, 77:977--1026, 2005.

\bibitem{Eijkel07}
J.~Eijkel.
\newblock Liquid slip in micro- and nanofluidics: recent research and its
  possible implications.
\newblock {\em Lab Chip}, 7:299--301, 2007.

\bibitem{Lauga07}
E.~Lauga, M.~P. Brenner, and H.~A. Stone.
\newblock Microfluidics: The no-slip boundary condition.
\newblock {\em Handbook of Experimental Fluid Dynamics, J. Foss, C. Tropea and
  A, Yarin (Hrsg), Springer, New York}, Chapter 15:cond--mat/0501557, 2008.

\bibitem{Bocquet07}
L.~Bocquet and J.~L. Barrat.
\newblock Flow boundary conditions from nano- to micro-scales.
\newblock {\em Soft Matter}, 3:685--693, 2007.

\bibitem{Navier1823}
C.~L. M.~H. Navier.
\newblock Sur les lois du mouvement des fluides.
\newblock {\em Mem. Acad. Roy. Sci. Inst. France}, 6:389--440, 1823.

\bibitem{Muller08}
M.~M{\"u}ller and C.~Pastorino.
\newblock Cyclic motion and inversion of surface flow direction in a dense
  polymer brush under shear.
\newblock {\em Europhys. Lett}, 81:28002, 2008.

\bibitem{Servantie08b}
J.~Servantie and M.~M{\"u}ller.
\newblock Temperature dependence of the slip length in polymer melts at
  attractive surfaces.
\newblock {\em Phys. Rev. Lett.}, 101:4989--5001, 2008.

\bibitem{Pastorino09}
C.~Pastorino, K.~Binder, and M.~M{\"u}ller.
\newblock Coarse-grained description of a brush-melt interface in equilibrium
  and under flow.
\newblock {\em Macromolecules}, 42:401--410, 2009.

\bibitem{Servantie08}
J.~Servantie and M.~M{\"u}ller.
\newblock Statics and dynamics of a cylindrical droplet under an external body
  force.
\newblock {\em J. Chem. Phys.}, 128:014709, 2008.

\bibitem{Kremer90}
K.~Kremer and G.~S. Grest.
\newblock Dynamics of entangled linear polymer melts - a molecular-dynamics
  simulation.
\newblock {\em J. Chem. Phys.}, 92:5057--5086, 1990.

\bibitem{LJ_pot2}
C.~Bennemann, W.~Paul, K.~Binder, and B.~D{\"u}nweg.
\newblock Molecular-dynamics simulations of the thermal glass transition in
  polymer melts: alpha-relaxation behavior.
\newblock {\em Phys. Rev. E}, 57:843--851, 1998.

\bibitem{Muller00i}
M.~M{\"u}ller and L.~G. Macdowell.
\newblock Interface and surface properties of short polymers in solution:
  {M}onte {C}arlo simulations and self-consistent field theory.
\newblock {\em Macromolecules}, 33:3902--3923, 2000.

\bibitem{Muller05}
M.~M{\"u}ller and G.~D. Smith.
\newblock Phase separation in binary mixtures containing polymers: a
  quantitative comparison of single-chain-in-mean-field simulations and
  computer simulations of the corresponding multichain systems.
\newblock {\em J. Polym. Sci. B: Polymer Physics}, 43:934--958, 2005.

\bibitem{Daoulas06b}
K.~Ch. Daoulas and M.~M{\"u}ller.
\newblock Single chain in mean field simulations: Quasi-instantaneous field
  approximation and quantitative comparison with {M}onte {C}arlo simulations.
\newblock {\em J. Chem. Phys.}, 125:184904, 2006.

\bibitem{Detcheverry09}
F.~A. Detcheverry, D.~Q. Pike, M.~M{\"u}ller, P.~F. Nealey, and J.~J. de~Pablo.
\newblock {M}onte {C}arlo simulation of coarse grain polymeric systems.
\newblock {\em Phys. Rev. Lett.}, 102:197801, 2009.

\bibitem{Kreer04}
T.~Kreer, S.~Metzger, M.~M{\"u}ller, K.~Binder, and J.~Baschnagel.
\newblock Static properties of end-tethered polymers in good solution: a
  comparison between different models.
\newblock {\em J. Chem. Phys.}, 120:4012--4023, 2004.

\bibitem{Grest99}
G.~S. Grest.
\newblock Normal and shear forces between polymer brushes.
\newblock {\em Adv. Polym. Sci}, 138:149--183, 1999.

\bibitem{DPD1}
P.~J. Hoogerbrugge and J.~M. V.~A. Koelman.
\newblock Simulating microscopic hydrodynamics phenomena with dissipative
  particle dynamics.
\newblock {\em Europhys. Lett}, 19:155, 1992.

\bibitem{DPD4b}
{P. Espanol }.
\newblock Hydrodynamics from dissipative particle dynamics.
\newblock {\em Phys. Rev. E}, 52:1734, 1995.

\bibitem{DPD2b}
P.~Warren and P.~Espanol.
\newblock Statistical-mechanics of dissipative particle dynamics.
\newblock {\em Europhys. Lett}, 30:191196, 1995.

\bibitem{DPD3b}
{T. Soddemann}, {B, Dnweg}, and {K. Kremer}.
\newblock Dissipative particle dynamics: a useful thermostat for equilibrium
  and nonequilibrium molecular dynamics simulations.
\newblock {\em Phys. Rev. E}, 68:46702, 2003.

\bibitem{Pastorino07}
C.~Pastorino, T.~Kreer, M.~M{\"u}ller, and K.~Binder.
\newblock Comparison of dissipative particle dynamics and langevin thermostats
  for out-of-equilibrium simulations of polymeric systems.
\newblock {\em Phys. Rev. E}, 76:026706, 2007.

\bibitem{PLIMPTON95}
S.J. Plimpton.
\newblock Fast parallel algorithms for short-range molecular dynamics.
\newblock {\em J. Comp. Phys.}, 117:1--19, 1995.

\bibitem{LAMMPS}
http://lammps.sandia.gov/.

\bibitem{Muller03d}
M.~M{\"u}ller, L.~G. MacDowell, and A.~Yethiraj.
\newblock Short chains at surfaces and interfaces: a quantitative comparison
  between density-functional theories and {M}onte {C}arlo simulations.
\newblock {\em J. Chem. Phys.}, 118:2929--2940, 2003.

\bibitem{Muller03c}
M.~M{\"u}ller and L.~G. MacDowell.
\newblock Wetting of polymer liquids: {M}onte {C}arlo simulations and
  self-consistent field calculations.
\newblock {\em J. Phys.: Condens. Matter}, 15:R609--R653, 2003.

\bibitem{Pastorino06}
C.~Pastorino, K.~Binder, T.~Kreer, and M.~M{\"u}ller.
\newblock Static and dynamic properties of the interface between a polymer
  brush and a melt of identical chains.
\newblock {\em J. Chem. Phys.}, 124:064902, 2006.

\bibitem{Macdowell06}
L.~G. MacDowell and M.~M{\"u}ller.
\newblock Adsorption of polymers on a brush: Tuning the order of the wetting
  phase transition.
\newblock {\em J. Chem. Phys.}, 124:084907, 2006.

\bibitem{MullerPlathe99}
F.~M{\"u}ller-Plathe.
\newblock Reversing the perturbation in nonequilibrium molecular dynamics: An
  easy way to calculate the shear viscosity of fluids.
\newblock {\em Phys. Rev. E}, 59:4894--4898, 1999.

\bibitem{Sikkenk1988}
M.~J.~P. Nijmeijer, A.~F. Bakker, C.~Bruin, and J.~H. Sikkenk.
\newblock A molecular dynamics simulation of the {L}ennard-{J}ones liquid-vapor
  interface.
\newblock {\em J. Chem. Phys.}, 89:3789, 1988.

\bibitem{Varnik00}
F.~Varnik, J.~Baschnagel, and K.~Binder.
\newblock Molecular dynamics results on the pressure tensor of polymer films.
\newblock {\em J. Chem. Phys.}, 113:4444--4453, 2000.

\bibitem{Bordat02}
P.~Bordat and F.~M{\"u}ller-Plathe.
\newblock The shear viscosity of molecular fluids: A calculation by reverse
  nonequilibrium molecular dynamics.
\newblock {\em J. Chem. Phys.}, 116:3362--3369, 2002.

\bibitem{Tenney10}
C.~M. Tenney and E.~J. Maginn.
\newblock Limitations and recommendations for the calculation of shear
  viscosity using reverse nonequilibrium molecular dynamics.
\newblock {\em J. Chem. Phys.}, 132, 2010.

\bibitem{hansen-mcdonald}
{J. P. Hansen} and {I. R. McDonald}.
\newblock {\em Theory of Simple Liquids}.
\newblock Academic Press, London, 1986.

\bibitem{Helfand75}
E.~Helfand.
\newblock Theory of inhomogeneous polymers: Fundamentals of the {G}aussian
  random--walk model.
\newblock {\em J. Chem. Phys.}, 62:999--1005, 1975.

\bibitem{Muller-Plathe02}
F.~M{\"u}ller-Plathe.
\newblock Coarse-graining in polymer simulation: From the atomistic to the
  mesoscopic scale and back.
\newblock {\em Chem. Phys. Chem.}, 3:754--769, 2002.

\bibitem{Reith03b}
D.~Reith, M.~Putz, and F.~M{\"u}ller-Plathe.
\newblock Deriving effective mesoscale potentials from atomistic simulations.
\newblock {\em J. Comp. Chem.}, 24:1624--1636, 2003.

\bibitem{book10}
M.~M{\"u}ller.
\newblock {\em Modeling and Simulations in Polymers, P.D. Gujrati and A.L.
  Leonov (edts)}, chapter Computational approaches for structure formation in
  multi-component polymer melts, pages 197--246.
\newblock Wiley-VCH Verlag, Weinheim, Germany, 2010.

\bibitem{Daoulas06}
K.~Ch. Daoulas, M.~M{\"u}ller, M.~P. Stoykovich, S.~M. Park, Y.~J.
  Papakonstantopoulos, J.~J. de~Pablo, P.~F. Nealey, and H.~H. Solak.
\newblock Fabrication of complex three-dimensional nanostructures from
  self-assembling block copolymer materials on two-dimensional chemically
  patterned templates with mismatched symmetry.
\newblock {\em Phys. Rev. Lett.}, 96:036104, 2006.

\bibitem{Daoulas06c}
K.~Ch. Daoulas, M.~M{\"u}ller, J.~J. de~Pablo, P.~F. Nealey, and G.~D. Smith.
\newblock Morphology of multi-component polymer systems: Single chain in mean
  field simulation studies.
\newblock {\em Soft Matter}, 2:573--583, 2006.

\bibitem{Doyle97b}
P.~S. Doyle, E.~S.~G. Shaqfeh, and A.~P. Gast.
\newblock Rheology of ''wet'' polymer brushes via brownian dynamics simulation:
  Steady vs oscillatory shear.
\newblock {\em Phys. Rev. Lett.}, 78:1182--1185, 1997.

\bibitem{Saphiannikova98}
M.~G. Saphiannikova, V.~A. Pryamitsyn, and T.~Cosgrove.
\newblock Self-consistent brownian dynamics simulation of polymer brushes under
  shear.
\newblock {\em Macromolecules}, 31:6662--6668, 1998.

\bibitem{Narayanan04}
B.~Narayanan, V.~A. Pryamitsyn, and V.~Ganesan.
\newblock Interfacial phenomena in polymer blends: a self-consistent brownian
  dynamics study.
\newblock {\em Macromolecules}, 37:10180--10194, 2004.

\bibitem{Rossky78}
P.~J. Rossky, J.~D. Doll, and H.~L. Friedman.
\newblock Brownian dynamics as smart {M}onte-{C}arlo simulation.
\newblock {\em J. Chem. Phys.}, 69:4628--4633, 1978.

\bibitem{allen86}
M.~Allen and D.~Tildesley.
\newblock {\em Computer Simulation of Liquids}.
\newblock Clarendon Press, Oxford, 1987.

\bibitem{MullerSL}
M.~M\"uller and K.~Ch. Daoulas.
\newblock Single-chain dynamics in a homogeneous melt and a lamellar
  microphase: A comparison between smart-{M}onte-{C}arlo dynamics,
  slithering-snake dynamics, and slip-link dynamics.
\newblock {\em J. Chem. Phys.}, 129:164906, 2008.

\bibitem{Doi}
M.~Doi and S.F. Edwards.
\newblock {\em The Theory of Polymer Dynamics}.
\newblock Oxford University Press, New York, 1994.

\bibitem{Wittmer04}
J.~P. Wittmer, H.~Meyer, J.~Baschnagel, A.~Johner, S.~Obukhov, L.~Mattioni,
  M.~M{\"u}ller, and A.~N. Semenov.
\newblock Long range bond-bond correlations in dense polymer solutions.
\newblock {\em Phys. Rev. Lett.}, 93:147801, 2004.

\bibitem{Narayanan06}
B.~Narayanan and V.~Ganesan.
\newblock Flow deformation of polymer blend droplets and the role of block
  copolymer compatibilizers.
\newblock {\em Phys. Fluids}, 18:042109, 2006.

\bibitem{Irving50}
J.~H. Irving and J.~G. Kirkwood.
\newblock The statistical mechanical theory of transport processes .4. the
  equations of hydrodynamics.
\newblock {\em J. Chem. Phys.}, 18:817--829, 1950.

\bibitem{Bocquet1993}
L.~Bocquet and J.-L. Barrat.
\newblock Hydrodynamic boundary conditions and correlation functions of
  confined fluids.
\newblock {\em Phys. Rev. Lett.}, 70:2726, 1993.

\bibitem{Bocquet1994}
L.~Bocquet and J.-L. Barrat.
\newblock Hydrodynamic boundary conditions, correlation functions, and kubo
  relations for confined fluids.
\newblock {\em Phys. Rev. E}, 49:3079, 1994.

\bibitem{Thompson90b}
P.~A. Thompson and M.~O. Robbins.
\newblock Shear-flow near solids - epitaxial order and flow
  boundary-conditions.
\newblock {\em Phys. Rev. A}, 41:6830--6837, 1990.

\bibitem{Bocquet94}
L.~Bocquet and J.~L. Barrat.
\newblock Hydrodynamic boundary-conditions, correlation-functions, and {K}ubo
  relations for confined fluids.
\newblock {\em Phys. Rev. E}, 49:3079--3092, 1994.

\bibitem{Cieplak01}
M.~Cieplak, J.~Koplik, and J.~R. Banavar.
\newblock Boundary conditions at a fluid-solid interface.
\newblock {\em Phys. Rev. Lett.}, 86:803--806, 2001.

\bibitem{Troian2004}
N.~V. Priezjev and S.~M. Troian.
\newblock Molecular origin and dynamic behavior of slip in sheared polymer
  films.
\newblock {\em Phys. Rev. Lett.}, 92:018302, 2004.

\bibitem{Priezjev09}
N.~V. Priezjev.
\newblock Shear rate threshold for the boundary slip in dense polymer films.
\newblock {\em Phys. Rev. E}, 80:031608, 2009.

\bibitem{Barrat99b}
J.~L. Barrat and L.~Bocquet.
\newblock Influence of wetting properties on hydrodynamic boundary conditions
  at a fluid/solid interface.
\newblock {\em Faraday Discussions}, 85:119--127, 1999.

\bibitem{Huang08c}
D.~M. Huang, C.~Sendner, D.~Horinek, R.~R. Netz, and L.~Bocquet.
\newblock Water slippage versus contact angle: A quasiuniversal relationship.
\newblock {\em Phys. Rev. Lett.}, 101, 2008.

\bibitem{Sendner09b}
C.~Sendner, D.~Horinek, L.~Bocquet, and R.~R. Netz.
\newblock Interfacial water at hydrophobic and hydrophilic surfaces: Slip,
  viscosity, and diffusion.
\newblock {\em Langmuir}, 25:10768--10781, 2009.

\bibitem{Muller09c}
M.~M{\"u}ller, C.~Pastorino, and J.~Servantie.
\newblock Hydrodynamic boundary condition of polymer melts at simple and
  complex surfaces.
\newblock {\em Comp. Phys. Comm.}, 180:600--604, 2009.

\bibitem{Alley83}
W.~E. Alley and B.~J. Alder.
\newblock Generalized transport-coefficients for hard-spheres.
\newblock {\em Phys. Rev. A}, 27:3158--3173, 1983.

\bibitem{Todd08}
B.~D. Todd, J.~S. Hansen, and P.~J. Daivis.
\newblock Nonlocal shear stress for homogeneous fluids.
\newblock {\em Phys. Rev. Lett.}, 100:195901, 2008.

\bibitem{Goel08}
G.~Goel, W.~P. Krekelberg, J.~R. Errington, and T.~M. Truskett.
\newblock Tuning density profiles and mobility of inhomogeneous fluids.
\newblock {\em Phys. Rev. Lett.}, 100:63--75, 2008.

\bibitem{Baschnagel05}
J.~Baschnagel and F.~Varnik.
\newblock Computer simulations of supercooled polymer melts in the bulk and
  in-confined geometry.
\newblock {\em J. Phys.: Condens. Matter}, 17:R851--R953, 2005.

\bibitem{Peter06}
S.~Peter, H.~Meyer, and J.~Baschnagel.
\newblock Thickness-dependent reduction of the glass-transition temperature in
  thin polymer films with a free surface.
\newblock {\em J. Polym. Sci. B: Polym. Phys.}, 44:2951--2967, 2006.

\bibitem{Fetzer05}
R.~Fetzer, K.~Jacobs, A.~M{\"u}nch, B.~Wagner, and T.~P. Witelski.
\newblock New slip regimes and the shape of dewetting thin liquid films.
\newblock {\em Phys. Rev. Lett.}, 95:127801, 2005.

\bibitem{Fetzer06}
R.~Fetzer, M.~Rauscher, A.~M{\"u}nch, B.~A. Wagner, and K.~Jacobs.
\newblock Slip-controlled thin-film dynamics.
\newblock {\em Europhys. Lett}, 75:638--644, 2006.

\bibitem{Fetzer07c}
R.~Fetzer, A.~M{\"u}nch, B.~Wagner, M.~Rauscher, and K.~Jacobs.
\newblock Hydrodynamic slip: A comprehensive analysis of dewetting profiles.
\newblock {\em Langmuir}, 23:10559--10566, 2007.

\bibitem{Fetzer07d}
R.~Fetzer and K.~Jacobs.
\newblock Slippage of newtonian liquids: Influence on the dynamics of dewetting
  thin films.
\newblock {\em Langmuir}, 23:11617--11622, 2007.

\bibitem{Muller01}
M.~M{\"u}ller, L.~G. MacDowell, P.~M{\"u}ller-Buschbaum, O.~Wunnike, and
  M.~Stamm.
\newblock Nano-dewetting: Interplay between van der {W}aals- and short-ranged
  interactions.
\newblock {\em J. Chem. Phys.}, 115:9960--9969, 2001.

\bibitem{Muller-Buschbaum05}
P.~M{\"u}ller-Buschbaum, O.~Wunnicke, M.~Stamm, Y.~C. Lin, and M.~M{\"u}ller.
\newblock Stability-instability transition by tuning the effective interface
  potential in polymeric bilayer films.
\newblock {\em Macromolecules}, 38:3406--3413, 2005.

\bibitem{Klein94b}
J.~Klein, E.~Kumacheva, D.~Mahalu, D.~Perahia, and L.~J. Fetters.
\newblock Reduction of frictional forces between solid-surfaces bearing polymer
  brushes.
\newblock {\em Nature}, 370:634--636, 1994.

\bibitem{Leibler94}
L.~Leibler, A.~Ajdari, A.~Mourran, G.~Coulon, and D.~Chatenay.
\newblock {\em Ordering in Macromolecular Systems, Proceedings of the Oums'93
  Toyonaka, Osaka, Japan, 3-6 June 1993 Springer-Verlag, Berlin}, 1994.

\bibitem{Macdowell05}
L.~G. MacDowell and M.~M{\"u}ller.
\newblock Observation of autophobic dewetting on polymer brushes from computer
  simulation.
\newblock {\em J. Phys.: Condens. Matter}, 17:S3523--S3528, 2005.

\bibitem{Muller00f}
M.~M{\"u}ller and L.~G. MacDowell.
\newblock Interface and surface properties of short polymers in solution:
  {M}onte {C}arlo simulations and self-consistent field theory.
\newblock {\em Macromolecules}, 33:3902--3923, 2000.

\bibitem{Muller01b}
M.~M{\"u}ller and L.~G. MacDowell.
\newblock Wetting of a short chain liquid on a brush: First-order and critical
  wetting transitions.
\newblock {\em Europhys. Lett}, 55:221--227, 2001.

\bibitem{Maas02}
J.~H. Maas, G.~J. Fleer, F.~A.~M. Leermakers, and M.~Cohen-Stuart.
\newblock Wetting of a polymer brush by a chemically identical polymer melt:
  Phase diagram and film stability.
\newblock {\em Langmuir}, 18:8871--, 2002.

\bibitem{voronov03}
A.~Voronov and O.~Shafranska.
\newblock Dependence of thin polystirene films stability on the thickness of
  grafted polystyrene brushes.
\newblock {\em Polymer}, 44:277--281, 2003.

\bibitem{voronov02}
A.~Voronov and O.~Shafranska.
\newblock Synthesis of chemically grafted polystyrene brushes and their
  influence on the dewetting in thin polystyrene films.
\newblock {\em Langmuir}, 18:4471--4477, 2002.

\bibitem{Dimitrov07}
D.~I. Dimitrov, A.~Milchev, and K.~Binder.
\newblock Polymer brushes in solvents of variable quality: Molecular dynamics
  simulations using explicit solvent.
\newblock {\em J. Chem. Phys.}, 127:084905, 2007.

\bibitem{Dimitrov08c}
D.~I. Dimitrov, A.~Milchev, and K.~Binder.
\newblock Local viscosity in the vicinity of a wall coated by polymer brush
  from green-kubo relations.
\newblock {\em Macromolecular Theory And Simulations}, 17:313--318, 2008.

\bibitem{Milner91b}
S.~T. Milner.
\newblock Hydrodynamic penetration into parabolic brushes.
\newblock {\em Macromolecules}, 24:3704--3705, 1991.

\bibitem{Brinkman47}
H.~C. Brinkman.
\newblock A calculation of the viscous force exerted by a flowing fluid on a
  dense swarm of particles.
\newblock {\em Appl. Sci. Res. A-Mechanics Heat Chemical}, 1:27--34, 1947.

\bibitem{Winkler06}
R.~G. Winkler.
\newblock Semiflexible polymers in shear flow.
\newblock {\em Phys. Rev. Lett.}, 97:128301, 2006.

\bibitem{Doyle00}
P.~S. Doyle, B.~Ladoux, and J.~L. Viovy.
\newblock Dynamics of a tethered polymer in shear flow.
\newblock {\em Phys. Rev. Lett.}, 84:4769--4772, 2000.

\bibitem{Gerashchenko06}
S.~Gerashchenko and V.~Steinberg.
\newblock Statistics of tumbling of a single polymer molecule in shear flow.
\newblock {\em Phys. Rev. Lett.}, 96:038304, 2006.

\bibitem{Gretz66}
R.~D. Gretz.
\newblock Line-tension effect in heterogeneous nucleation.
\newblock {\em Surface Science}, 5:239, 1966.

\bibitem{Gretz66b}
R.~D. Gretz.
\newblock Line-tension effect in a surface energy model of a cap-shaped
  condensed phase.
\newblock {\em J. Chem. Phys.}, 45:3160, 1966.

\bibitem{MacDowell02b}
L.~G. MacDowell, M.~M{\"u}ller, and K.~Binder.
\newblock How do droplets on a surface depend on the system size?
\newblock {\em Colloids Surf. A}, 206:277--291, 2002.

\bibitem{Neumanna1998}
A.~Amirfazlia, D.~Y. Kwoka, J.~Gaydosb, and A.~W. Neumanna.
\newblock Line tension measurements through drop size dependence of contact
  angle.
\newblock {\em J. Colloid Int. Sci.}, 205:1, 1998.

\bibitem{Kulmala2007}
A.~I. Hienola, P.~M. Winkler, P.~E. Wagner, H.~Vehkamki, A.~Lauri, I.~Napari,
  and M.~Kulmala.
\newblock Estimation of line tension and contact angle from heterogeneous
  nucleation experimental data.
\newblock {\em J. Chem. Phys.}, 126:094705, 2007.

\bibitem{Fredrickson92b}
G.~H. Fredrickson, A.~Ajdari, L.~Leibler, and J.~P. Carton.
\newblock Surface-modes and deformation energy of a molten polymer brush.
\newblock {\em Macromolecules}, 25:2882--2889, 1992.

\bibitem{Xi96}
H.~W. Xi and S.~T. Milner.
\newblock Surface waves on polymer brushes.
\newblock {\em Macromolecules}, 29:4772--4776, 1996.

\bibitem{Likhtman99}
A.~E. Likhtman, S.~H. Anastasiadis, and A.~N. Semenov.
\newblock Theory of surface deformations of polymer brushes in solution.
\newblock {\em Macromolecules}, 32:3474--3480, 1999.

\bibitem{Pomeau1999}
L.~Mahadevan and Y.~Pomeau.
\newblock Rolling droplets.
\newblock {\em Phys. Fluids}, 11:2449, 1999.

\bibitem{Pomeau2001}
U.~Thiele, M.~G. Velarde, K.~Neuffer, M.~Bestehorn, and Y.~Pomeau.
\newblock Sliding drops in the diffuse interface model coupled to
  hydrodynamics.
\newblock {\em Phys. Rev. E}, 64:061601, 2001.

\bibitem{Thiele02}
U.~Thiele, K.~Neuffer, M.~Bestehorn, Y.~Pomeau, and M.~G. Velarde.
\newblock Sliding drops on an inclined plane.
\newblock {\em Colloids and Surfaces A-Physicochemicaland Engineering Aspects},
  206:87--104, 2002.

\bibitem{Mognetti10}
B.~M. Mognetti, H.~Kusumaatmaja, and J.~M. Yeommans.
\newblock Drop dynamics on hydrophobic and superhydrophobic surfaces.
\newblock {\em Faraday Discuss.}, 146:in press, 2010.

\bibitem{Huppert82}
H.~E. Huppert.
\newblock Flowand instability of a viscous current down a slope.
\newblock {\em Nature}, 300:427--429, 1982.

\bibitem{Rio05}
E.~Rio, A.~Daerr, B.~Andreotti, and L.~Limat.
\newblock Boundary conditions in the vicinity of a dynamic contact line:
  Experimental investigation of viscous drops sliding down an inclined plane.
\newblock {\em Phys. Rev. Lett.}, 94:024503, 2005.

\bibitem{LeGrand05}
N.~Le Grand, A.~Daerr, and L.~Limat.
\newblock Shapeand motion of drops sliding down an inclined plane.
\newblock {\em J. Fluid Mechanics}, 541:293--315, 2005.

\bibitem{LeGrand06b}
N.~Le Grand-Piteira, A.~Daerr, and L.~Limat.
\newblock Meandering rivulets on a plane: a simple balance between inertiaand
  capillarity.
\newblock {\em Phys. Rev. Lett.}, 96:254503, 2006.

\bibitem{Kim02d}
H.~Y. Kim, H.~J. Lee, and B.~H. Kang.
\newblock Sliding of liquid drops down an inclined solid surface.
\newblock {\em J. Coll. Interf. Sci.}, 247:372--380, 2002.

\bibitem{BrochardF1992}
F.~Brochard-Wyart and P.~G. de~Gennes.
\newblock Dynamics of partial wetting.
\newblock {\em Adv. Coll. Interf. Sci.}, 39:1--11, 1992.

\bibitem{Tanner79}
L.~H. Tanner.
\newblock Spreading of silicone oil drops on horizontal surfaces.
\newblock {\em J. Phys. D}, 12:1473, 1979.

\bibitem{Ruijter97}
M.~J. de~Ruijter, J.~de~Coninck, T.~D. Blake, A.~Clarke, and A.~Rankin.
\newblock Contact angle relaxation during the spreading of partially wetting
  drops.
\newblock {\em Langmuir}, 13:7293--7298, 1997.

\bibitem{Ruijter99}
M.~J. de~Ruijter, T.~D. Blake, and J.~de~Coninck.
\newblock Dynamic wetting studied by molecular modeling simulations of droplet
  spreading.
\newblock {\em Langmuir}, 15:7836--7847, 1999.

\bibitem{Milchev02e}
A.~Milchev and K.~Binder.
\newblock Droplet spreading: a {M}onte {C}arlo test of tanner's law.
\newblock {\em J. Chem. Phys.}, 116:7691--7694, 2002.

\bibitem{Heine03}
D.~R. Heine, G.~S. Grest, and E.~B. Webb.
\newblock Spreading dynamics of polymer nanodroplets.
\newblock {\em Phys. Rev. E}, 68:061603, 2003.

\bibitem{Heine04}
D.~R. Heine, G.~S. Grest, and E.~B. Webb.
\newblock Spreading dynamics of polymer nanodroplets in cylindrical geometries.
\newblock {\em Phys. Rev. E}, 70:011606, 2004.

\bibitem{Webb2005}
D.~R. Heine, G.~S. Grest, and E.~B. Webb.
\newblock Surface wetting of liquid nanodroplets: Droplet-size effects.
\newblock {\em Phys. Rev. Lett.}, 95:107801, 2005.

\bibitem{Heine05b}
D.~R. Heine, G.~S. Grest, and E.~B. Webb.
\newblock Diverse spreading behavior of binary polymer nanodroplets.
\newblock {\em Langmuir}, 21:7959--7963, 2005.

\bibitem{Blake69}
T.~D. Blake and J.~M. Haynes.
\newblock Kinetics of liquid/liquid displacement.
\newblock {\em J. Coll. Interf. Sci.}, 30:421, 1969.

\bibitem{Bertrand07}
E.~Bertrand, T.~D. Blake, V.~Ledauphin, G.~Ogonowski, and J.~de~Coninck.
\newblock Dynamics of dewetting at the nanoscale using molecular dynamics.
\newblock {\em Langmuir}, 23:3774--3785, 2007.

\bibitem{Voue07}
M.~Voue, R.~Rioboo, M.~H. Adao, J.~Conti, A.~I. Bondar, D.~A. Ivanov, T.~D.
  Blake, and J.~de~Coninck.
\newblock Contact-line friction of liquid drops on self-assembled monolayers:
  Chain-length effects.
\newblock {\em Langmuir}, 23:4695--4699, 2007.

\bibitem{Shanahan88}
M.~E.~R. Shanahan.
\newblock The spreading dynamics of a liquid-drop on a viscoelastic solid.
\newblock {\em J. Phys. D}, 21:981--985, 1988.

\bibitem{Carre93}
A.~Carre and M.~E.~R. Shanahan.
\newblock Viscoelastic braking of drop spreading.
\newblock {\em Comptes Rendus De L Academie Des Sciences Serie II},
  317:1153--1158, 1993.

\bibitem{Shanahan94}
M.~E.~R. Shanahan and A.~Carre.
\newblock Anomalous spreading of liquid-drops on an elastomeric surface.
\newblock {\em Langmuir}, 10:1647--1649, 1994.

\bibitem{Shanahan95}
M.~E.~R. Shanahan and A.~Carre.
\newblock Viscoelastic dissipation in wetting and adhesion phenomena.
\newblock {\em Langmuir}, 11:1396--1402, 1995.

\bibitem{Carre96}
A.~Carre, J.~C. Gastel, and M.~E.~R. Shanahan.
\newblock Viscoelastic effects in the spreading of liquids.
\newblock {\em Nature}, 379:432--434, 1996.

\bibitem{Xie98}
R.~Xie, A.~Karim, J.~F. Douglas, C.~C. Han, and R.~A. Weiss.
\newblock Spinodal dewetting of thin polymer films.
\newblock {\em Phys. Rev. Lett.}, 81:1251--1254, 1998.

\bibitem{Brochardwyart94b}
F.~Brochard-Wyart, P.~G. de~Gennes, H.~Hervert, and C.~Redon.
\newblock Wetting and slippage of polymer melts on semi-ideal surfaces.
\newblock {\em Langmuir}, 10:1566--1572, 1994.

\bibitem{Reiter92}
G.~Reiter.
\newblock Dewetting of thin polymer-films.
\newblock {\em Phys. Rev. Lett.}, 68:75--78, 1992.

\bibitem{Reiter93}
G.~Reiter.
\newblock Unstable thin polymer-films - rupture and dewetting processes.
\newblock {\em Langmuir}, 9:1344--1351, 1993.

\bibitem{seemann01b}
R.~Seemann, S.~Herminghaus, and K.~Jacobs.
\newblock Dewetting patterns and molecular forces: a reconciliation.
\newblock {\em Phys. Rev. Lett.}, 86:5534--5537, 2001.

\bibitem{Reiter01b}
G.~Reiter.
\newblock Dewetting of highly elastic thin polymer films.
\newblock {\em Phys. Rev. Lett.}, 87, 2001.

\bibitem{Herminghaus02}
S.~Herminghaus, R.~Seemann, and K.~Jacobs.
\newblock Generic morphologies of viscoelastic dewetting fronts.
\newblock {\em Phys. Rev. Lett.}, 89:056101, 2002.

\bibitem{Rauscher05}
M.~Rauscher, A.~M{\"u}nch, B.~Wagner, and R.~Blossey.
\newblock A thin-film equation for viscoelastic liquids of jeffreys type.
\newblock {\em Eur. Phys. J. E}, 17:373--379, 2005.

\bibitem{Munch06}
A.~M{\"u}nch, B.~Wagner, M.~Rauscher, and R.~Blossey.
\newblock A thin-film model for corotational jeffreys fluids under strong slip.
\newblock {\em Eur. Phys. J. E}, 20:365--368, 2006.

\bibitem{Blossey06}
R.~Blossey, A.~M{\"u}nch, M.~Rauscher, and B.~Wagner.
\newblock Slip vs. viscoelasticity in dewetting thin films.
\newblock {\em Eur. Phys. J. E}, 20:267--271, 2006.

\bibitem{Baumchen2009}
O.~B{\"a}umchen, R.~Fetzer, and K.~Jacobs.
\newblock Reduced interfacial entanglement density affects the boundary
  conditions of polymer flow.
\newblock {\em Phys. Rev. Lett.}, 103:247801, 2009.

\bibitem{Homberg08}
M.~H{\"o}mberg and M.~M{\"u}ller.
\newblock Generating multichain configurations of an inhomogeneous melt from
  the knowledge of single-chain properties.
\newblock {\em J. Chem. Phys.}, 128:1198--1211, 2008.

\bibitem{Fetzer07b}
R.~Fetzer, M.~Rauscher, R.~Seemann, K.~Jacobs, and K.~Mecke.
\newblock Thermal noise influences fluid flow in thin films during spinodal
  dewetting.
\newblock {\em Phys. Rev. Lett.}, 99, 2007.

\bibitem{IZVEKOV05_2}
S.~Izvekov and G.A. Voth.
\newblock Multiscale coarse graining of liquid-state systems.
\newblock {\em J. Chem. Phys.}, 123:134105, 2005.

\bibitem{Praprotnik05}
M.~Praprotnik, L.~delle Site, and K.~Kremer.
\newblock Adaptive resolution molecular-dynamics simulation: Changing the
  degrees of freedom on the fly.
\newblock {\em J. Chem. Phys.}, 123:224106, 2005.

\bibitem{Praprotnik08}
M.~Praprotnik, L.~delle Site, and K.~Kremer.
\newblock Multiscale simulation of soft matter: From scale bridging to adaptive
  resolution.
\newblock {\em Annual Review Of Physical Chemistry}, 59:545--571, 2008.

\bibitem{Lu10}
L.~Y. Lu, S.~Izvekov, A.~Das, H.~C. Andersen, and G.~A. Voth.
\newblock Efficient, regularized, and scalable algorithms for multiscale
  coarse-graining.
\newblock {\em Journal Of Chemical Theory And Computation}, 6:954--965, 2010.

\bibitem{Delgado-Buscalioni03b}
R.~Delgado-Buscalioni and P.~V. Coveney.
\newblock Continuum-particle hybrid coupling for mass, momentum, and energy
  transfers in unsteady fluid flow.
\newblock {\em Phys. Rev. E}, 67:046704, 2003.

\bibitem{Fabritis06}
G.~de~Fabritiis, R.~Delgado-Buscalioni, and P.~V. Coveney.
\newblock Multiscale modeling of liquids with molecular specificity.
\newblock {\em Phys. Rev. Lett.}, 97:134501, 2006.

\bibitem{Koumoutsakos05}
P.~Koumoutsakos.
\newblock Multiscale flow simulations using particles.
\newblock {\em Annual Rev. Fluid Mech.}, 37:457--487, 2005.

\bibitem{Delgado-Buscalioni08}
R.~Delgado-Buscalioni, K.~Kremer, and M.~Praprotnik.
\newblock Concurrent triple-scale simulation of molecular liquids.
\newblock {\em J. Chem. Phys.}, 128:114110, 2008.

\bibitem{Oron97}
A.~Oron, S.~H. Davis, and S.~G. Bankoff.
\newblock Long-scale evolution of thin liquid films.
\newblock {\em Rev. Mod. Phys.}, 69:931--980, 1997.

\end{thebibliography}

\end{document}